\pdfoutput=1

\documentclass[11pt,twoside,a4paper,cmspaper,final,collab]{cms-tdr}
\def\svnVersion{65026:65688}\def\cmsCernNo{2011-099}\def\cmsCernDate{\today}\def\cmsMessage{Submitted to Physical Review D}

\begin{document}\cmsNoteHeader{SUS-10-009}

\hyphenation{had-ron-i-za-tion}
\hyphenation{cal-or-i-me-ter}
\hyphenation{de-vices}
\RCS$Revision: 65688 $
\RCS$HeadURL: svn+ssh://alverson@svn.cern.ch/reps/tdr2/papers/SUS-10-009/trunk/SUS-10-009.tex $
\RCS$Id: SUS-10-009.tex 65688 2011-07-06 13:15:25Z alverson $
\providecommand{\PSGc}{\widetilde{\chi}}
\providecommand{\PSGcz}{\widetilde{\chi}^0}
\cmsNoteHeader{SUS-10-009}
\title{\texorpdfstring{Inclusive search for squarks and gluinos \\  in $\Pp\Pp$ collisions at $\sqrt{s}=7\TeV$}{Inclusive search for squarks and gluinos in pp collisions at sqrt(s)=7 TeV}}
\author{The CMS Collaboration}
\date{\today}

\abstract{A search is performed for heavy particle pairs produced in
  $\sqrt{s}$ = 7\TeV proton-proton collisions with 35\pbinv of
  data collected by the CMS experiment at the LHC.  The search is
  sensitive to squarks and gluinos of  generic supersymmetry models, provided
  they are kinematically accessible, with minimal assumptions on
  properties of the lightest superpartner particle.  The kinematic
  consistency of the selected events is tested against the hypothesis
  of heavy particle pair production using the dimensionless \textit{razor}
  variable $R$, related to the missing transverse energy
  \ETm. The new physics signal is characterized by a broad
  peak in the distribution of $M_R$, an event-by-event indicator of
  the heavy particle mass scale. This new approach is complementary to
  \ETm-based searches.  After background modeling based on
  data, and background rejection based on $R$ and $M_R$, no significant
  excess of events is found beyond the standard model
  expectations. The results are interpreted in the context of the
  constrained minimal supersymmetric standard model as well as two
  simplified supersymmetry models. }

\hypersetup{%
pdfauthor={CMS Collaboration},%
pdftitle={Inclusive search for squarks and gluinos in pp collisions at sqrt(s) = 7 TeV},%
pdfsubject={CMS},%
pdfkeywords={CMS, LHC, SUSY, Razor, top+X, W+jets, Z+jets, QCD scaling}}

\maketitle 
\newcommand{\hilight}[1]{\colorbox{yellow}{#1}}

\section{Introduction}

Models with softly broken supersymmetry
(SUSY)~\cite{Ramond,Golfand,Volkov,Wess,Fayet} predict superpartners
of the standard model (SM) particles.  Experimental limits from the
Tevatron and LEP showed that superpartner particles, if they exist,
are significantly heavier than their SM counterparts.  Proposed
experimental searches for $R$-parity conserving SUSY at the Large
Hadron Collider (LHC) have therefore focused on a combination of two
SUSY signatures: multiple energetic jets and/or leptons from the
decays of pair-produced squarks and gluinos, and large missing
transverse energy (\ETm) from the two weakly interacting
lightest superpartners (LSP) produced in separate decay chains.

In this article a new approach is presented that is inclusive not only
for SUSY but also in the larger context of physics beyond the standard
model. The focal point for this novel \textit{razor}
analysis~\cite{rogan} is the production of pairs of heavy particles
(of which squarks and gluinos are examples), whose masses are
significantly larger than those of any SM particle. The analysis is
designed to kinematically discriminate the pair production of heavy
particles from SM backgrounds, without making strong assumptions about
the \ETm spectrum or details of the decay chains of these
particles. The baseline selection requires two or more reconstructed
objects, which can be calorimetric jets, isolated electrons or
isolated muons. These objects are grouped into two
  \textit{megajets}. The razor analysis tests the consistency, event by
event, of the hypothesis that the two megajets represent the visible
portion of the decays of two heavy particles.  This strategy is
complementary to traditional searches for signals in the tails of the
\ETm distribution ~\cite{:2007ww, Aaltonen:2008rv, Collaboration:2011xk,
  daCosta:2011qk, Aad:2011hh, Aad:2011xm, RA2, alphaT, :2011wb, Chatrchyan:2011bz}
and is applied to data collected with the Compact Muon Solenoid (CMS)
detector from pp collisions at $\sqrt{s}=7\TeV$ corresponding to an
integrated luminosity of 35\pbinv.

\section{The CMS Apparatus}

A description of the CMS detector can be found
elsewhere~\cite{:2008zzk}.
A characteristic feature of the CMS detector is its superconducting
solenoid magnet, of 6~m internal diameter, providing a field of
3.8~T. The silicon pixel and strip tracker, the crystal
electromagnetic calorimeter (ECAL) and the brass/scintillator hadron
calorimeter (HCAL) are contained within the solenoid. Muons are
detected in gas-ionization chambers embedded in the steel return
yoke. The ECAL has an energy resolution of better than 0.5\,\% above
100\GeV. The HCAL combined with the ECAL, measures the jet energy with
a resolution $\Delta E/E \approx 100\,\% / \sqrt{E/\GeV} \oplus
5\,\%$.

CMS uses a coordinate system with the origin located at the nominal
collision point, the $x$-axis pointing towards the center of the LHC,
the $y$-axis pointing up (perpendicular to the LHC plane), and the
$z$-axis along the counterclockwise beam direction. The azimuthal
angle $\phi$ is measured with respect to the $x$-axis in the $xy$
plane and the polar angle $\theta$ is defined with respect to the
$z$-axis. The pseudorapidity is $\eta = -\ln [\tan(\theta / 2)]$.

\section{The Razor Analysis\label{intro}}

The pair production of two heavy particles, each decaying to an unseen
LSP plus jets, gives rise to a generic SUSY-like signal.  Events in
this analysis are forced into a dijet topology by combining all jets
in the event into two megajets.
When an isolated lepton is present, it can be included
in the megajets or not, as described in Sections~\ref{sec:es}
and~\ref{sec:be}.  To the extent that the pair of megajets accurately
reconstructs the visible portion of the underlying parent particle
decays, the kinematic properties of the signal are equivalent to the pair production of,
for example, two heavy squarks $\PSq_1$, $\PSq_2$, with
$\PSq_i\rightarrow {{\rm{j}}_i} \PSGcz_i $, for $i=1,~2$,
where  ${\rm{j}}_i$ and $\PSGcz_i$  denote the visible and invisible
products of the decays, respectively. In the approximation that the
heavy squarks are produced at threshold and  their visible decay
products are massless, the center of mass (CM) frame four-momenta are

\begin{eqnarray}
&&\hspace*{-50pt}
p_{\rm j_1}  =  \frac{M_\Delta}{2}(1,\hat{\boldmath{u}}_1)  \;,\quad
p_{\rm j_2}  =  \frac{M_\Delta}{2}(1,\hat{\boldmath{u}}_2)\; , \\
&&\hspace*{-50pt}
p_{\PSGc_{1}} =  \frac{M_\Delta}{2} \left( \frac{2M_{\PSq}}{M_\Delta}-1,-\hat{\boldmath{u}}_1\right) \; ,\quad
p_{\PSGc_{2}}  =  \frac{M_\Delta}{2} \left(
  \frac{2M_{\PSq}}{M_\Delta}-1,-\hat{\boldmath{u}}_2 \right) \; ,
\end{eqnarray}

where $\hat{\boldmath{u}}_i$ is the unit vector in the direction of $\rm{j}_i$, and

\begin{eqnarray}
&&\hspace*{-50pt}
M_\Delta \equiv  \frac{M_{\PSq}^{2}-M_{\PSGc}^{2}}{M_{\PSq}}~,~ \;
\end{eqnarray}

where $M_{\PSq}$ and $M_{\PSGc}$ are the squark and LSP
masses, respectively.

In events with two undetected particles in the partonic final state, it
is not possible to reconstruct the actual CM frame. Instead, an
approximate event-by-event  reconstruction is made assuming the dijet signal
topology, replacing the CM frame with the  \textit{$R$ frame} \cite{rogan},
defined as the longitudinally boosted frame that equalizes the magnitude
of the two megajets' three-momenta. The $R$ frame would be the CM frame for
signal events, if the squarks were produced at threshold and if the CM
system had no overall transverse momentum from initial-state
radiation.  The longitudinal Lorentz boost factor is defined by
\begin{equation}
\beta_{R}\equiv\frac{E^{\rm j_1}-E^{\rm j_2}}{p^{\rm j_1}_{z}-p^{\rm j_2}_{z}}\; ,
\end{equation}
where $E^{\rm j_1}$, $E^{\rm j_2}$ and $p^{\rm j_1}_{z}$, $p^{\rm
  j_2}_{z}$ are the megajet energies and longitudinal momenta ,
respectively.  To the extent that the $R$ frame matches the true CM
frame, the maximum value of the scalar sum of the megajets'
transverse momenta ($\PT^1,~\PT^2$) is $M_\Delta$ for signal
events. The maximum value of the \ETm is also $M_\Delta$.  A
transverse mass $M_T^{R}$ is defined whose maximum value for signal events is
also $M_\Delta$ in the limit that the $R$ and CM frames coincide:

\begin{equation}
M_T^{R}\equiv \sqrt{ \frac{\ETm(\PT^{\rm j_1}+\PT^{\rm j_2}) -
    \VEtmiss {\mathbf \cdot}
    (\vec{p}_T^{\,\rm j_1}+\vec{p}_T^{\,\rm j_2})}{2}} \; .
\end{equation}

The event-by-event estimator of $M_\Delta$ is
\begin{equation}
M_R\equiv 2|\vec{p}^{R}_{{\rm j_1}}| = 2|\vec{p}^{R}_{{\rm j_2}}|\; ,
\end{equation}
where $\vec{p}^{R}_{{\rm j_1}}$ and $\vec{p}^{R}_{{\rm j_2}}$ are the
3-momenta of the megajets in the $R$ frame.  For signal events in the
limit where the $R$ frame and the true CM frame coincide, $M_R$ equals
$M_\Delta$, and more generally $M_R$ is expected to peak around
$M_\Delta$ for signal events.
For QCD dijet and multijet events the only relevant scale is
$\sqrt{\hat{s}}$, the CM energy of the partonic subprocess.  The
search for an excess of signal events in a tail of a distribution is
thus recast as a search for a peak on top of a steeply falling SM
residual tail in the $M_R$ distribution. To extract the peaking
signal, the QCD multijet background needs to be reduced to manageable
levels. This is achieved using the \textit{razor} variable defined as:
\begin{equation}
  R\equiv \frac{M_T^{R}}{M_R} \; .\end{equation}

\begin{figure*}[ht!]
\begin{center}
\includegraphics[width=0.49\textwidth]{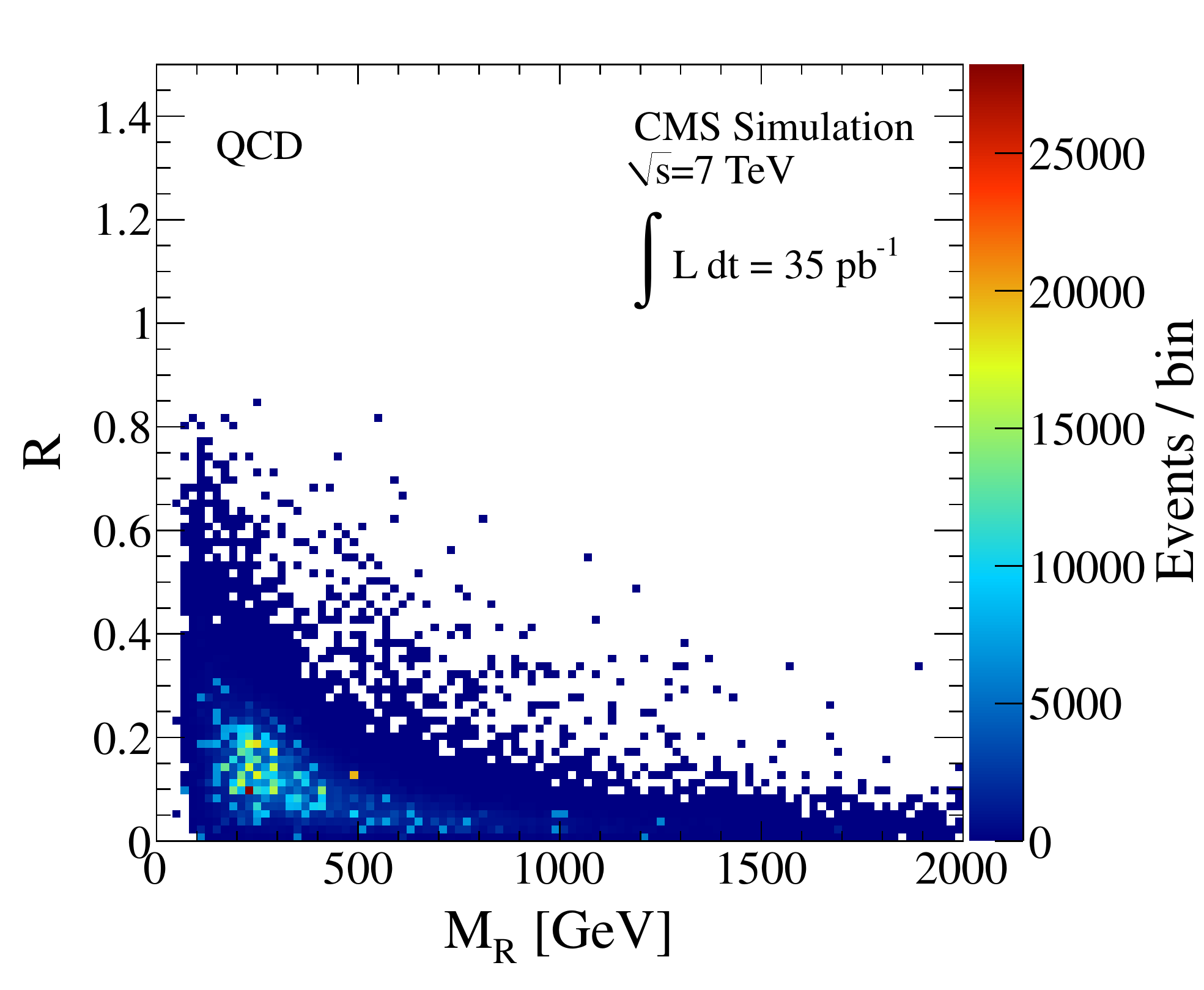}
\includegraphics[width=0.49\textwidth]{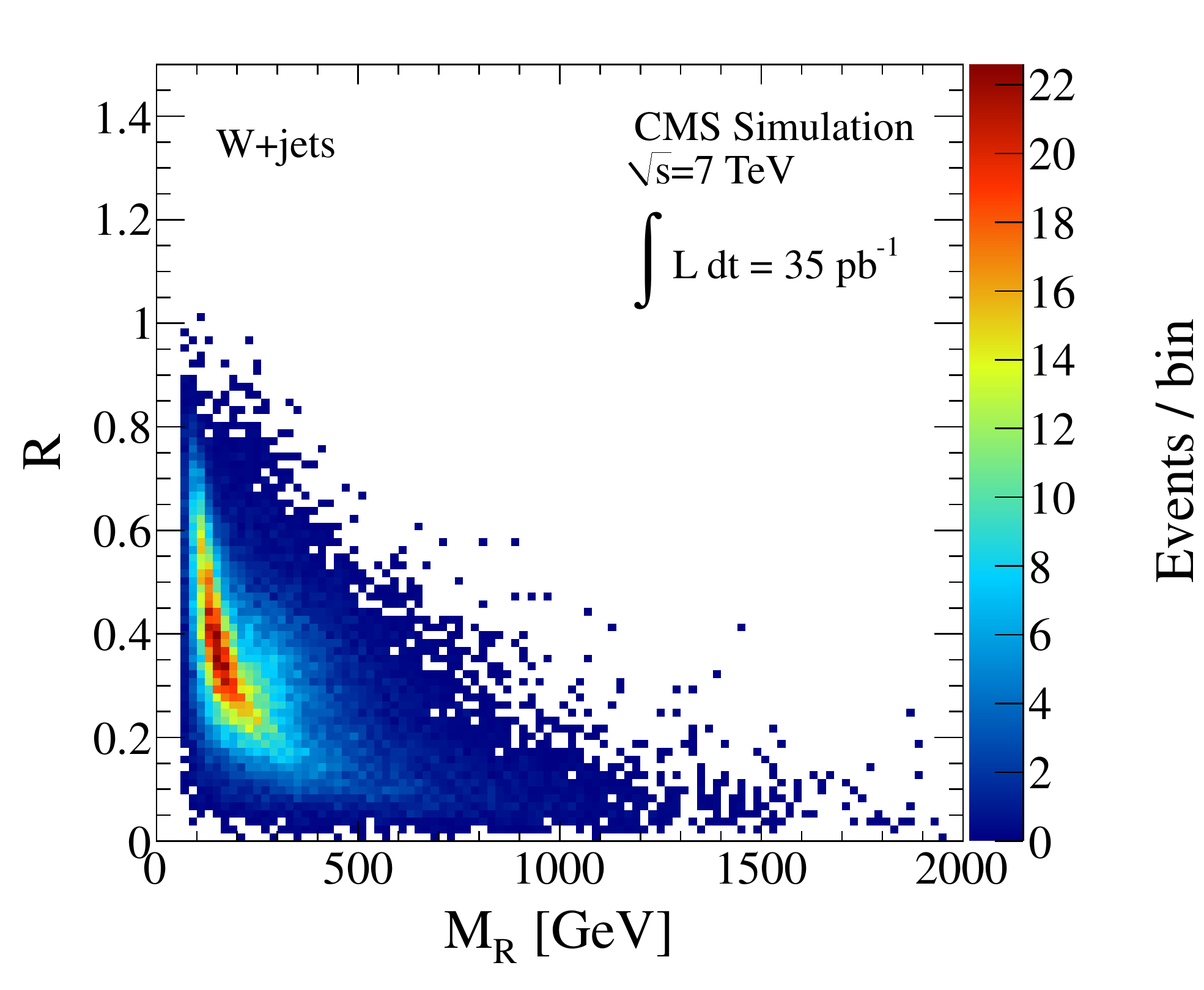}
\includegraphics[width=0.49\textwidth]{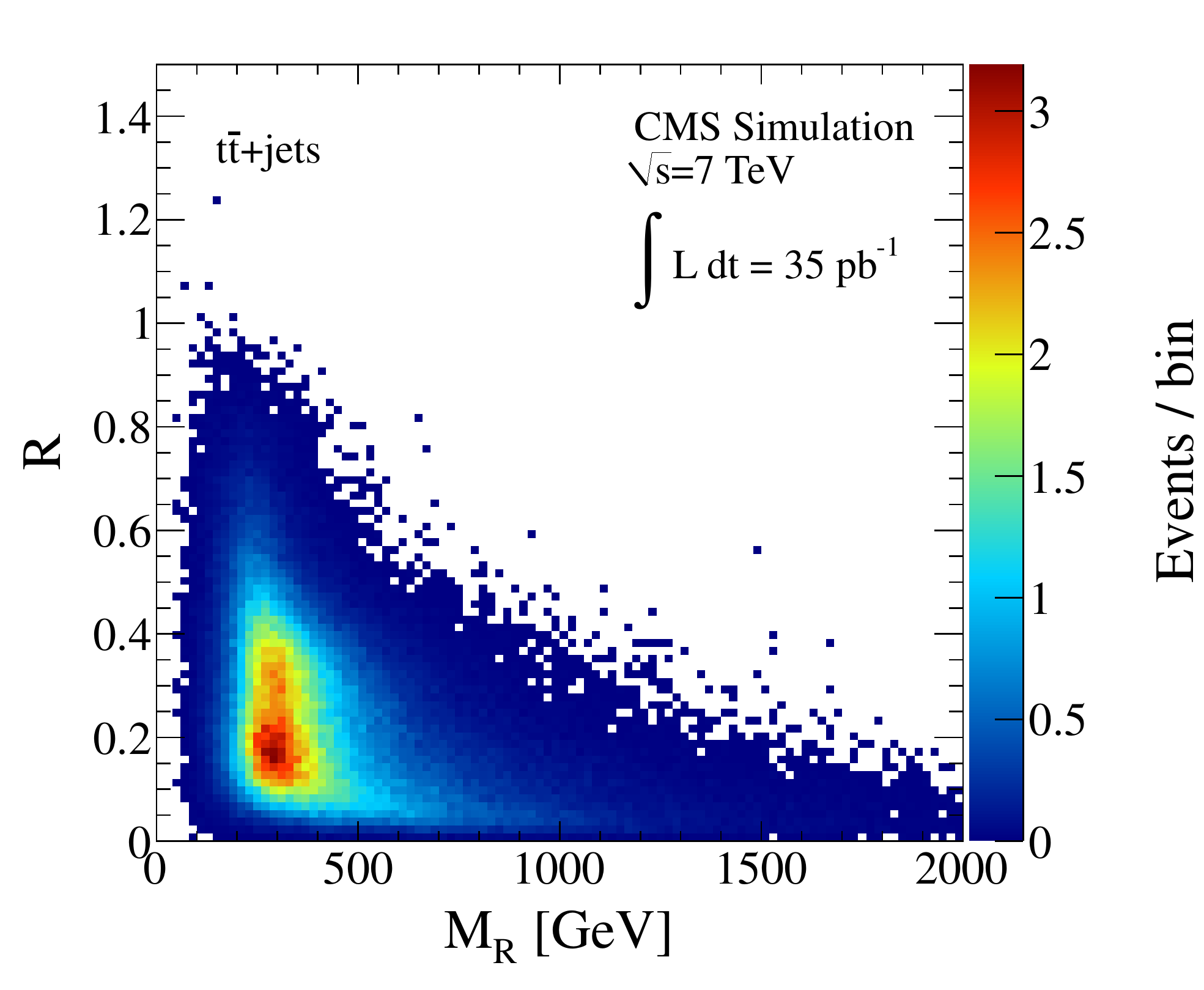}
\includegraphics[width=0.49\textwidth]{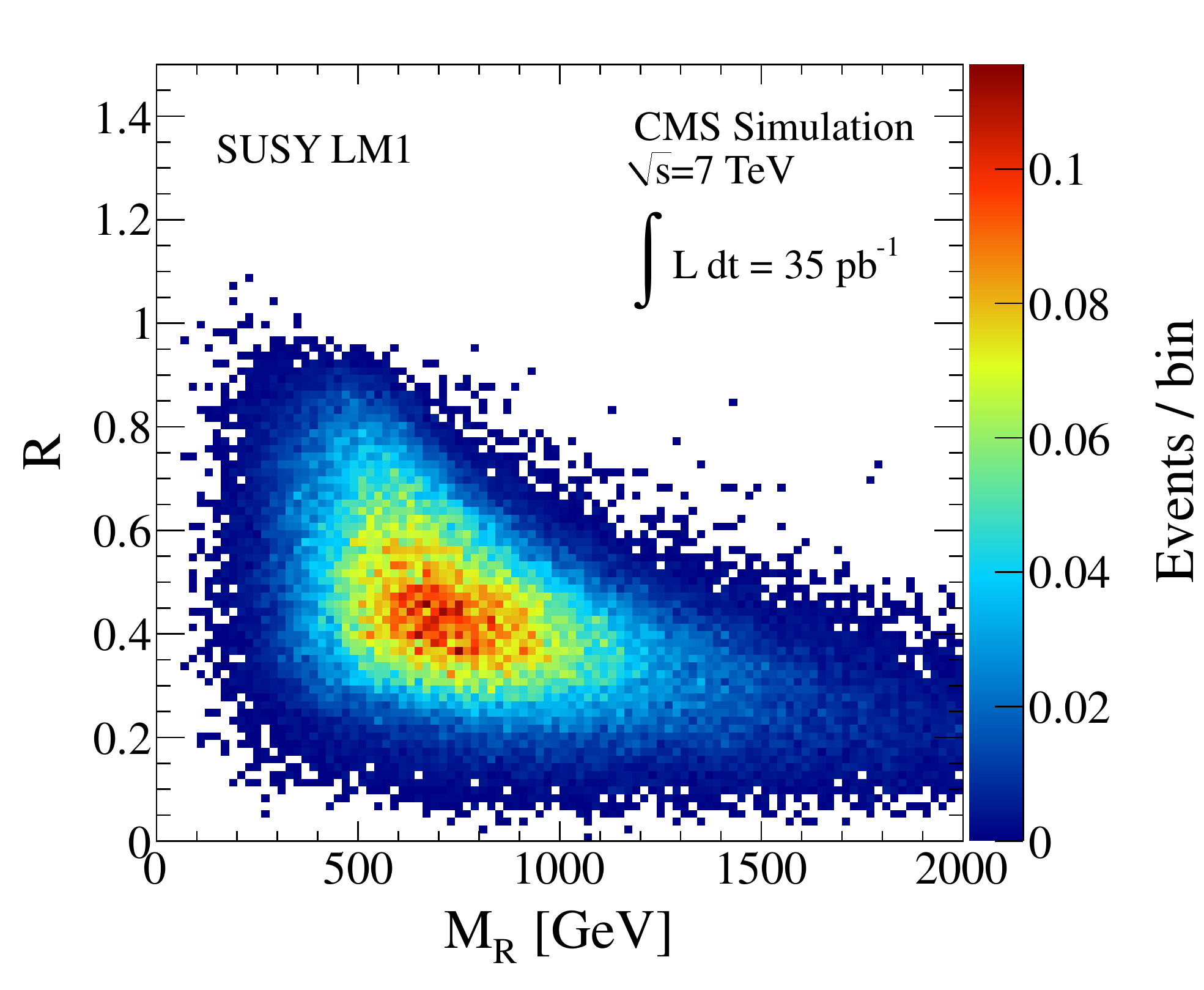}
\caption{ Scatter plot in the ($M_R$, $R$) plane for simulated events:
  (top left) QCD multijet, (top right) $\PW$+jets, (bottom left)
  $\ttbar$+jets, and (bottom right) the SUSY benchmark model LM1
  \cite{PTDR2} with $M_\Delta = 597\GeV$. The yields are normalized to
  an integrated luminosity of 35\pbinv. The bin size is (20\GeV
  $\times$ 0.015). \label{fig:MR_v_R}}
\end{center}
\end{figure*}

Since for signal events $M_T^R$ has a maximum value of $M_\Delta$
(i.e., a kinematic edge), $R$ has a maximum value of approximately 1
and the distribution of $R$ for signal events peaks around 0.5. These
properties motivate the appropriate kinematic requirements for the
signal selection and background reduction. It is noted that, while $M_T^R$
and $M_R$ measure the same scale (one as an end-point, the other as a
peak), they are largely uncorrelated for signal events, as shown in
Fig.~\ref{fig:MR_v_R}. In this figure, the $\PW$+jets and $\ttbar$+jets
backgrounds peak at $M_R$ values partially determined by the $\PW$ and top quark masses,
respectively.

In this analysis the SM background shapes and normalizations are
obtained from data. The backgrounds are extracted from control regions
in the $R$ and $M_R$ distributions dominated by SM processes.
Initial estimates of the background distributions in these regions are
obtained from the individual simulated background components, but
their  shapes  and normalizations are then corrected using data.
The analysis flow is as follows:

\begin{enumerate}
\item The inclusive data sets are collected using
  the electron, muon,  and hadronic-jet  triggers.
\item These data sets are examined for the presence of a
  well-identified isolated electron or muon, irrespective of the
  trigger path. Based on the presence or absence of such a lepton,
  each event is assigned to one of three disjoint event samples,
  referred to as the electron, muon, and hadronic \textit{boxes}.  These
  boxes serve as controls of processes in the SM with leptons, jets,
  and neutrinos, e.g. QCD multijet, $\PW$+jets or $\cPZ$+jets, and
  $\cPqt$+X. The diboson background is found to be negligible.  Exclusive
  multilepton boxes are also defined but are not sufficiently
  populated to be used in this analysis.
\item Megajets are constructed for events passing a baseline kinematic
  selection, and the $R$ and $M_R$ event variables are computed. In
  the electron box, electrons are clustered with jets in the
  definition of the megajets. Jets matched to these electrons are
  removed to avoid double-counting. In the muon box, muons are
  included in the megajet clustering.
\item In order to characterize the distribution of the SM background
  events in the ($M_R$, $R$) plane, a kinematic region is identified
  in the lepton boxes that is dominated by $\PW(\ell \nu)+$jets. Another
  region is found that is dominated by the sum of the non-QCD
  backgrounds.
\item Events remaining in the hadronic box primarily consist of QCD
  multijet, $\cPZ(\nu\bar{\nu})$+jets, $\PW(\ell\nu)$+jets, and $\cPqt$+X
  events that produce $\ell$+jets+$\ETm$ final states with
  charged leptons that do not satisfy the electron or muon selections.
  The shapes and normalizations of these non-QCD background processes
  in the hadronic box are estimated using the results from the lepton
  boxes in appropriate regions in the ($M_R$, $R$) plane.
\item The QCD background shape and normalization in each of the lepton
  boxes is extracted by reversing the lepton isolation requirements to
  obtain control samples dominated by QCD background.
\item The QCD background in the hadronic box is estimated using QCD
  control samples collected with prescaled jet triggers.
\item The large-$R$ and high-$M_R$ regions of all boxes are signal candidate
  regions not used for the background estimates.  Above a given
  $R$ threshold, the $M_R$ distribution of the backgrounds observed in the
  data is well modeled by simple exponential functions.  Having determined the $R$
  and $M_R$ shape and normalization of the backgrounds in the control
  regions, the SM yields are extrapolated to the large-$R$
  and high-$M_R$ signal candidate regions for each box.
\end{enumerate}

\section{Event Selection\label{sec:es}}

The analysis uses data sets recorded with triggers based on the
presence of an electron, a muon, or on $H_T$, the uncorrected scalar
sum of the transverse energy of jets reconstructed at the trigger
level.  Prescaled jet triggers with low thresholds are used  for the
QCD multijet background estimation in the hadronic box.

The  analysis is guided by studies of Monte Carlo  (MC)  event
samples generated with the {\sc Pythia} \cite{Sjostrand:2006za} and
{\sc Madgraph} \cite{Maltoni:2002qb} programs, simulated using the CMS
{\sc Geant}-based \cite{G4} detector simulation, and then processed by
the same software used to reconstruct real collision data.  Events
with QCD multijet, top quarks, and electroweak bosons were generated
with {\sc Madgraph} interfaced with {\sc Pythia} for parton showering,
hadronization, and underlying event description. To generate Monte
Carlo samples for SUSY, the mass spectrum was first calculated with
{\sc {Softsusy}} \cite{softsusy} and the decays with {\sc {Susyhit}}
\cite{Susyhit}. The {\sc {Pythia}} program was used with the {\sc SLHA}
interface \cite{SLHA} to generate the events. The generator level
cross section and the K factors for the next-to-leading order  (NLO) cross
section calculation were computed using {\sc Prospino} \cite{prospino}.

Events are required to have at least one good reconstructed
interaction vertex \cite{TRK-10-005}. When multiple vertices are
found, the one with the highest associated $\sum_{\rm track}\PT$ is
used.  Jets are reconstructed offline from calorimeter energy deposits
using the infrared-safe anti-k$_{\rm{T}}$~\cite{antikt} algorithm with
radius parameter $0.5$. Jets are corrected for the nonuniformity of
the calorimeter response in energy and $\eta$ using corrections
derived with the simulation and are required to have $\PT> 30\GeV$
and $|\eta| < 3.0$.  The jet energy scale uncertainty for these
corrected jets is $5\%$ \cite{JES}. The \ETm is
reconstructed using the particle flow algorithm \cite{PFMET}.

The electron and muon reconstruction and identification criteria are
described in ~\cite{EWK-PAS}. Isolated electrons and muons are
required to have $\PT>20\GeV$ and $\vert\eta|<$ 2.5 and 2.1,
respectively, and to satisfy the selection requirements from
\cite{EWK-PAS}.  The typical lepton trigger and reconstruction
efficiencies are 98\% and 99\%, respectively, for electrons and 95\%
and 98\% for muons.

The reconstructed hadronic jets, isolated electrons, and isolated
muons are grouped into two megajets, when at least two such objects
are present in the event.  The megajets are constructed as a sum of
the four-momenta of their constituent objects.  After considering all
possible partitions of the objects into two megajets, the combination
minimizing the invariant masses summed in quadrature of the resulting
megajets is selected among all combinations for which the $R$ frame is
well defined.

After the construction of the two megajets the boost variable
$|\beta_{R}|$ is computed; due to the approximations mentioned above,
$|\beta_{R}|$ can fall in an unphysical region (${\ge}1$) for signal or
background events; these events are removed. The additional
requirement $|\beta_{R}|\le 0.99$ is imposed to remove events for
which the razor variables become singular. This requirement is typically
85\%    efficient for simulated SUSY events. The azimuthal angular
difference between the megajets is required to be less than $2.8$ radians;
this requirement suppresses nearly back-to-back QCD dijet
events. These requirements define the inclusive baseline
selection. After this selection, the signal efficiency in the
constrained minimal supersymmetric standard model
(CMSSM)~\cite{Chamseddine,Barbieri,Hall,Kane} parameter space for a gluino
mass of  ${\sim} 600\GeV$ is over 50\%.

\section{Background Estimation\label{sec:be}}

In traditional searches for SUSY based on missing transverse energy, it
is difficult to model the tails of the \ETm
distribution and the contribution from events with spurious
instrumental effects. The QCD multijet production is an especially
daunting background because of its very high cross section and complicated
modeling of its high-\PT and \ETm tails. In this analysis
a cut on $R$ makes it possible to isolate the QCD multijet
background in the low-$M_R$ region.

Apart from QCD multijet backgrounds, the remaining backgrounds in the
lepton and hadronic boxes are processes with genuine \ETm due to
energetic neutrinos and leptons from massive vector boson decays
(including $\PW$ bosons from top quark decays). After applying an $R$ threshold,
the $M_{R}$ distributions in the lepton and hadronic boxes are
very similar for these backgrounds; this similarity is exploited in
the modeling and normalization of these backgrounds.

\subsection{QCD multijet background\label{sec:qcd}}

The QCD multijet control sample for the hadronic box is defined from
event samples recorded with prescaled jet triggers and passing the
baseline analysis selection for events without a well-identified
isolated electron or muon. The trigger requires at least two jets with
an average uncorrected $\PT > 15\GeV$.  Because of the low jet
threshold, the QCD multijet background dominates this sample for low
$M_R$, thus allowing the extraction of the $M_R$ shapes with different
$R$ thresholds for QCD multijet events. These shapes are corrected
for the $H_T$ trigger turn-on efficiency.

\begin{figure*}[htbp]
\begin{center}
\includegraphics[width=0.47\textwidth]{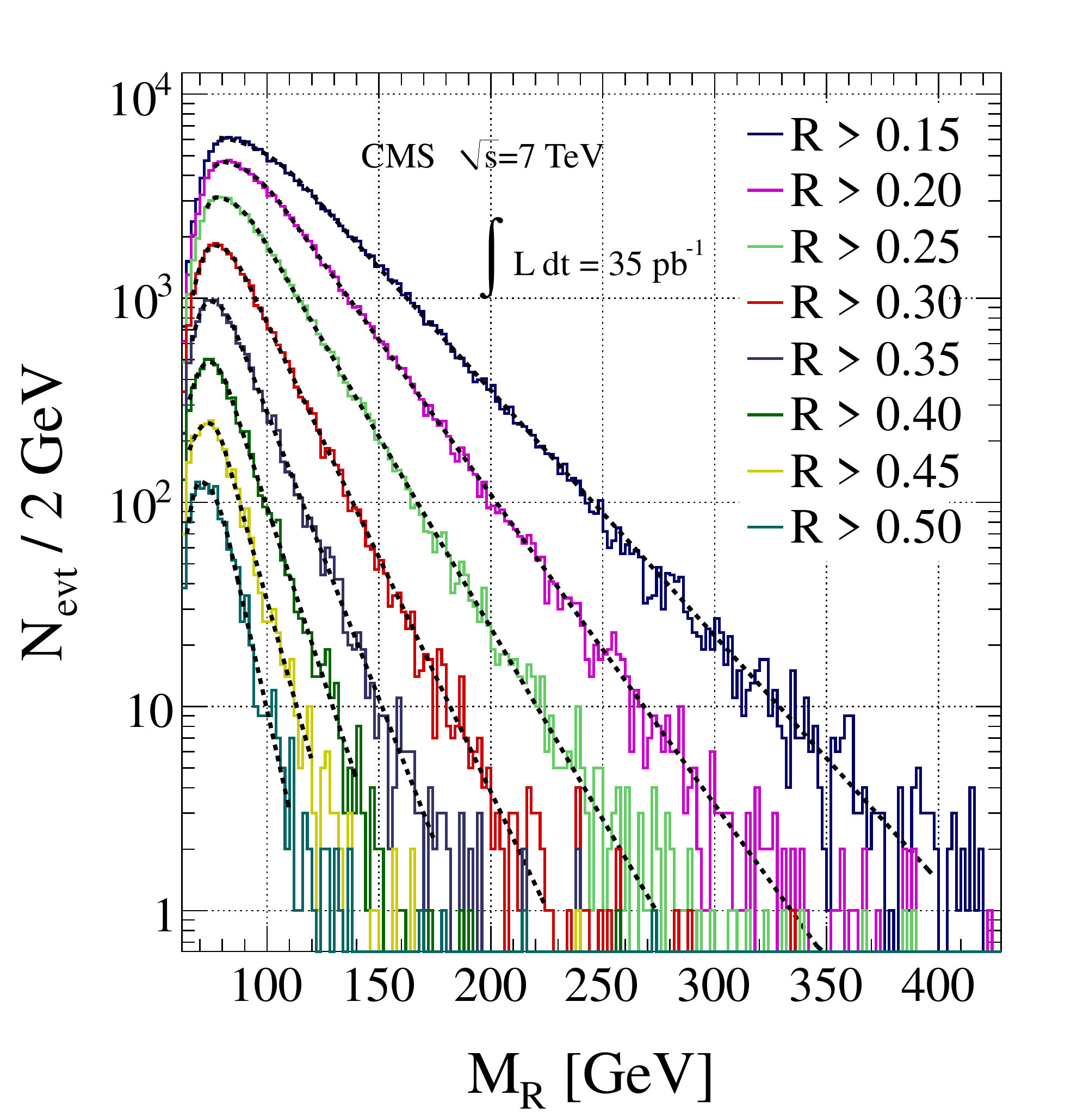}
\includegraphics[width=0.47\textwidth]{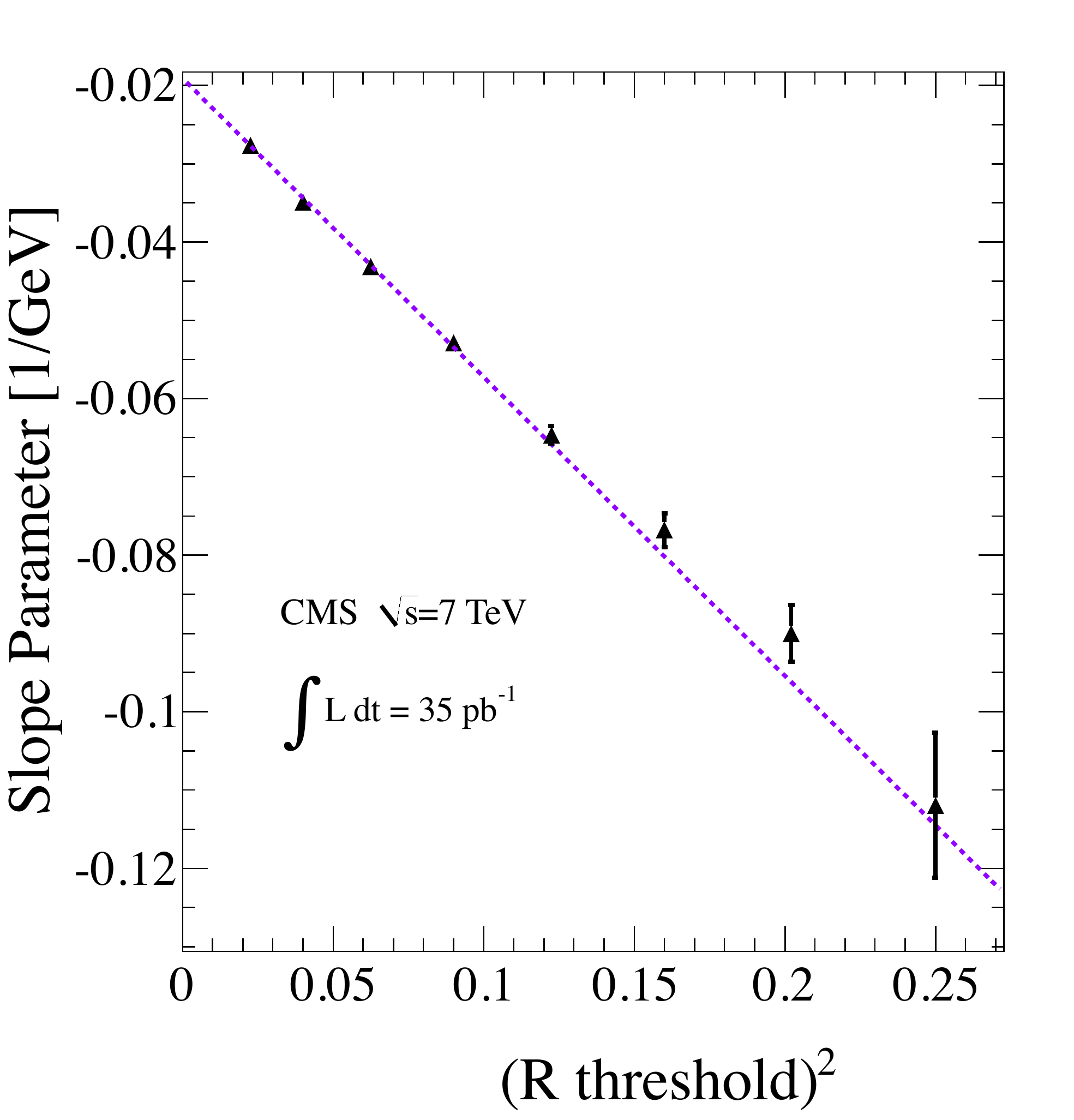}
\caption{(Left) $M_{R}$ distributions for different values of the $R$
  threshold for data events in the QCD control box. Fits of the
  $M_{R}$ distribution to an exponential function and an asymmetric
  Gaussian at low $M_R$, are shown as dotted black curves.  (Right)
  The exponential slope $S$ from fits to the $M_{R}$ distribution, as
  a function of the square of the $R$ threshold for data events in the
  QCD control box.}
\label{fig:DATA_QCD_calo}
\end{center}
\end{figure*}

The $M_{R}$ distributions for events satisfying the QCD control box
selection, for different values of the $R$ threshold, are shown in
Fig.~\ref{fig:DATA_QCD_calo} (left). The $M_{R}$ distribution is
exponentially falling, after a turn-on at low $M_{R}$ resulting from
the \PT threshold requirement on the jets entering the megajet
calculation. After the turn-on which is fitted with an asymmetric
Gaussian, the exponential region of these distributions is fitted for
each value of $R$ to extract the exponential slope, denoted by $S$.
The value of $S$ that maximizes the likelihood in the exponential fit
is found to be a linear function of $R^{2}$, as shown in
Fig.~\ref{fig:DATA_QCD_calo} (right); fitting $S$ to the form $S = a +
bR^{2}$ determines the values of $a$ and $b$.

When measuring the exponential slopes of the $M_{R}$ distributions as
a function of the $R$ threshold, the correlations due to events
satisfying multiple $R$ threshold requirements are neglected.  The
effect of these correlations on the measurement of the slopes is
studied by using pseudo-experiments and is found to be negligible.

To measure the shape of the QCD background component in the lepton boxes, the corresponding
lepton trigger data sets are used with the baseline selection and
reversed lepton isolation criteria. The QCD background component in the lepton
boxes is found to be negligible.

The $R$ threshold shapes the $M_R$ distribution in a simple therefore
predictable way. Event selections with combined $R$ and $M_R$
thresholds are found to suppress jet mismeasurements, including
severe mismeasurements of the electromagnetic or hadronic
component of the jet energy, or other anomalous calorimetric noise
signals such as the ones described in~\cite{hcalnoise,ecalnoise}.

\subsection{\texorpdfstring{\PW+jets, \cPZ+jets, and \cPqt+X backgrounds}{W+jets, Z+jets, and t+X backgrounds}\label{bg-prop}}

Using the muon (MU) and electron (ELE) control boxes defined in
Section~\ref{intro}, $M_R$ intervals dominated by $\PW(\ell\nu)$+jets
events are identified for different $R$ thresholds.  In both simulated
and data events, the $M_{R}$ distribution is well described by two
independent exponential components. The first component of
$W(\ell\nu)$+jets corresponds to events where the highest \PT
object in one of the megajets is the isolated electron or muon;
the second component consists of events where the leading object
in both megajets is a jet, as is typical also for the $t$+X background
events. The first component of $W(\ell\nu)$+jets can be measured
directly in data, because it dominates over all other backgrounds in a control region
of lower $M_R$ set by  the $R$ threshold.
At higher values of $M_R$, the
first component of $W(\ell\nu)$+jets falls off rapidly,
and the remaining background is instead dominated by the
sum of  $t$+X and the second component of $W(\ell\nu)$+jets;
this defines a second control region of intermediate $M_R$
set by the $R$ threshold.

\begin{figure*}[htbp]
\begin{center}
\includegraphics[width=0.49\textwidth]{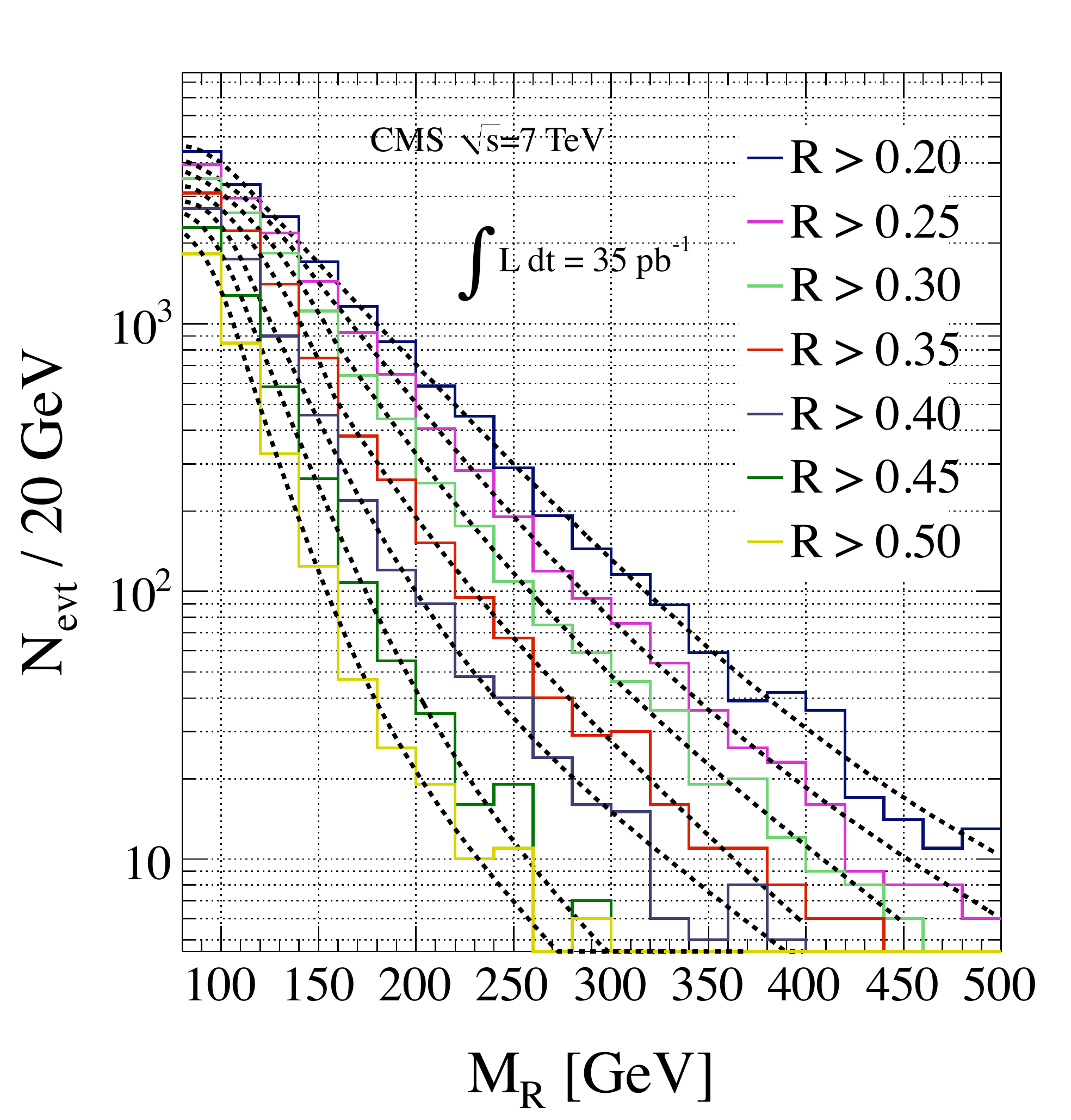}
\includegraphics[width=0.49\textwidth]{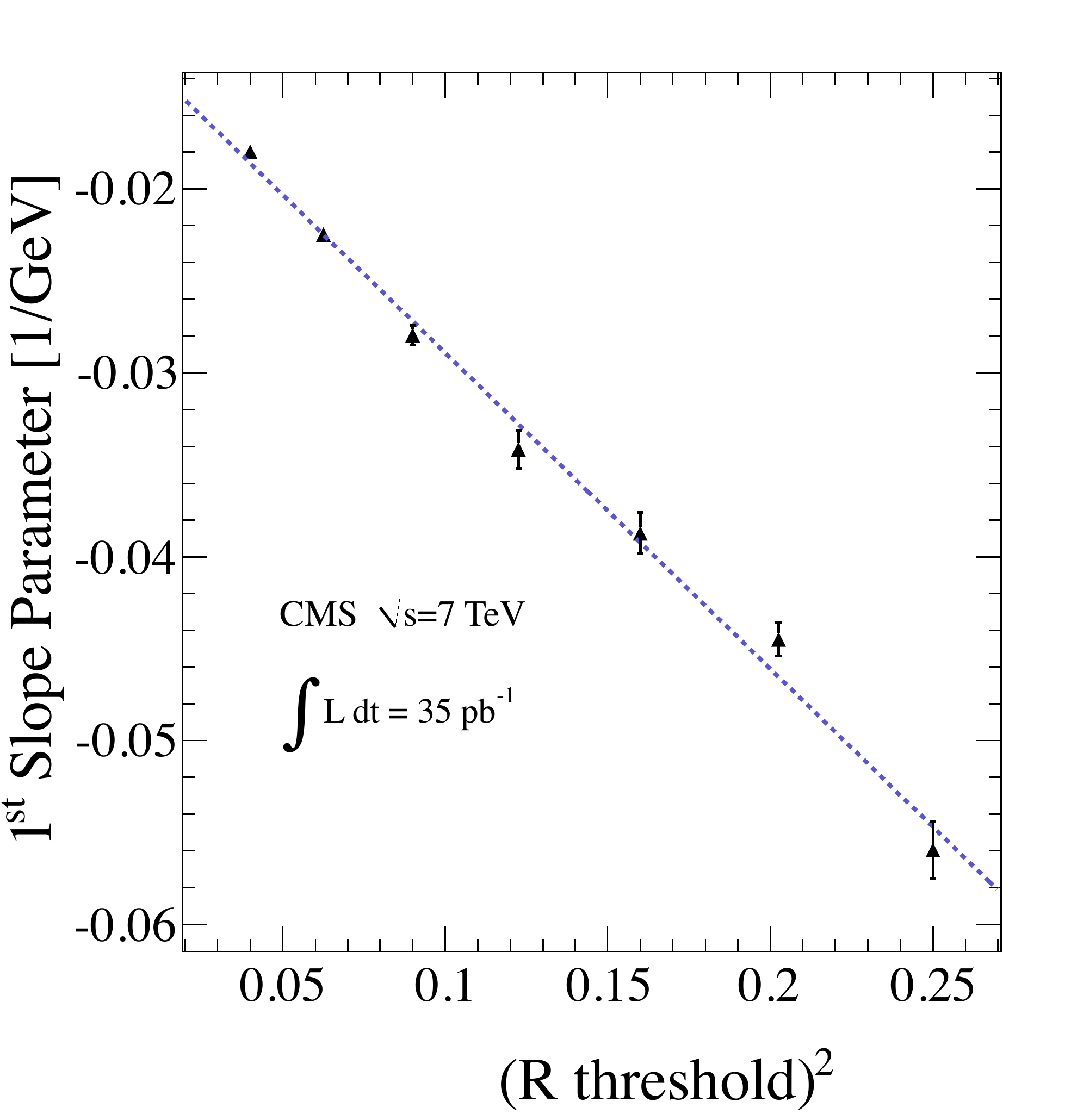}\\
\includegraphics[width=0.49\textwidth]{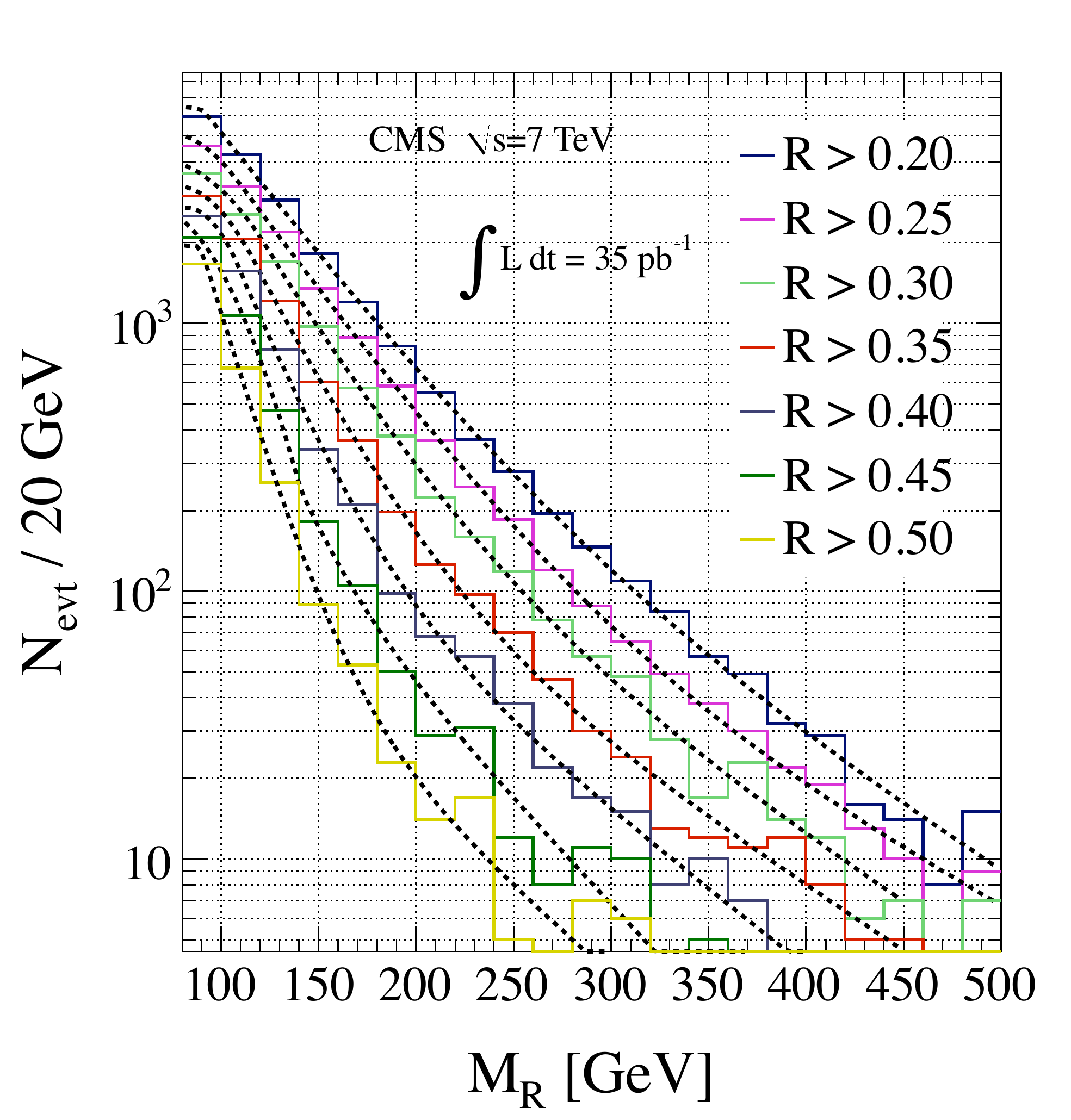}
\includegraphics[width=0.49\textwidth]{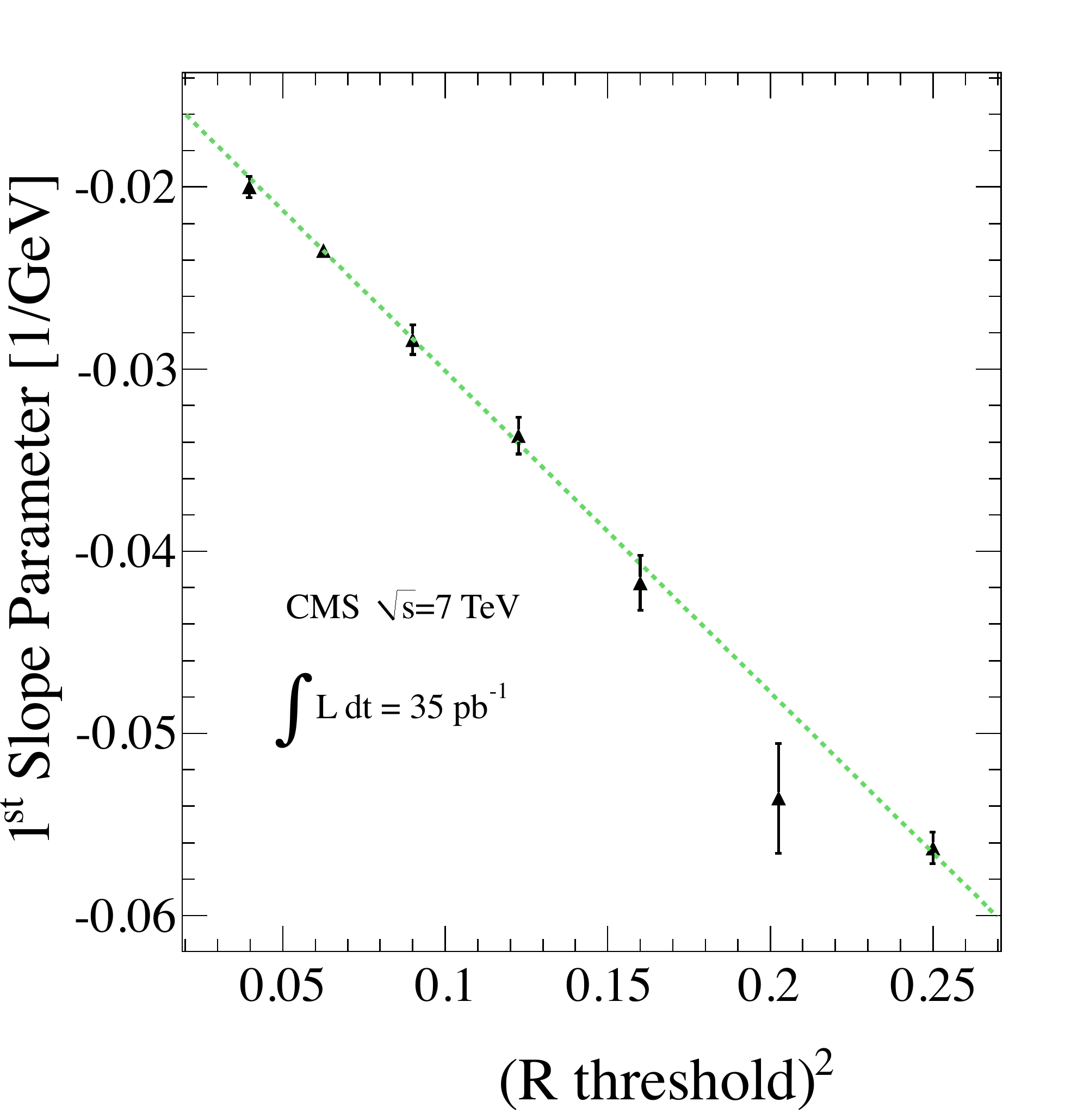}
\caption{(Left) $M_{R}$ distributions for different values of the $R$
  threshold from data events selected in the MU (upper) and ELE
  (lower) boxes. Dotted curves show the results of fits using two
  independent exponential functions and an asymmetric Gaussian at low
  $M_R$. (Right) The slope $S$ of the first exponential component
  as a function of the square of the $R$ threshold in the MU (upper)
  and ELE (lower) boxes. The dotted lines show the results of the fits
  to the form $S = a + bR^{2}$.}
\label{fig:DATA_MU_slopes}
\end{center}
\end{figure*}

Using these two control regions in a given box,
a simultaneous fit determines both exponential slopes along
with the absolute normalization of the first component of $W(\ell\nu)$+jets
and the relative normalization of the sum of the second
component of $W(\ell\nu)$+jets  with the other backgrounds.
The $M_R$ distributions as a function of $R$ are shown in
Fig.~\ref{fig:DATA_MU_slopes} (left).  The slope parameters
characterizing the exponential behavior of the first
$W(\ell\nu)$+jets component are shown in Fig.~\ref{fig:DATA_MU_slopes}
(right); they are consistent within uncertainties between the electron
and muon channels. The values of the parameters $a$ and $b$ that
describe the $R^{2}$ dependence of the slope are in good agreement
with the values extracted from simulated $W(\ell\nu)$+jets events.

The  data/MC  ratios $\rho(a)_1^{\mathrm{data/MC}}$,  $\rho(b)_1^{\mathrm{data/MC}}$
of the  first component slope parameters $a$, $b$  measured in the MU and ELE boxes are thus combined
yielding

\begin{eqnarray}
\rho(a)_{1}^{\mathrm{data/MC}} = 0.97 \pm 0.02 ~~;~~
\rho(b)_{1}^{\mathrm{data/MC}} = 0.97 \pm 0.02~~,
\end{eqnarray}
where the quoted uncertainties are determined from the fits.

The ratios
$\rho^{\mathrm{data/MC}}$
are taken as correction factors for the shapes of the
$Z$+jets and $t+$X backgrounds as extracted from simulated samples
for the MU and ELE boxes; the same corrections are used
for the shape of the first component of $W(\ell\nu)$+jets as
extracted from simulated samples for the
hadronic (HAD) box.

The data/MC correction factors for
the $\cPZ(\nu\bar\nu)$+jets and $\cPqt$+X backgrounds in the HAD box, as well as
the second component of $\PW(\ell\nu)$+jets in the MU, ELE, and HAD boxes,
are measured in the  MU and ELE boxes
using a \textit{lepton-as-neutrino} treatment of leptonic events. Here
the electron or muon is excluded from the megajet reconstruction,
kinematically mimicking the presence of an additional neutrino.  With
the lepton-as-neutrino treatment in the MU and ELE boxes only one
exponential component is observed both in data and in
$\PW(\ell\nu)$+jets simulated events.  In the simulation, the value of
this single exponential component slope is found to agree with the
value for the second component of $\PW(\ell\nu)$+jets obtained in
the default treatment.

The combined data/MC correction factors measured using this
lepton-as-neutrino treatment are
\begin{eqnarray}
\rho(a)_{2}^{\mathrm{data/MC}} = 1.01 \pm 0.02 ~~;~~
\rho(b)_{2}^{\mathrm{data/MC}} = 0.94 \pm 0.07.
\end{eqnarray}
For the final background prediction the magnitude of the relative
normalization between the two $\PW(\ell\nu)$+jets components, denoted
$f^{\PW}$,  is determined from a binned maximum likelihood fit
in the region 200 $< M_{R} < $ 400\GeV.

\section{Results}
\subsection{Lepton box background predictions \label{sec:LEPBOX}}

Having extracted the $M_{R}$ shape of the $\PW$+jets and $\cPZ$+jets
backgrounds, their relative normalization is set from the $\PW$ and $\cPZ$
cross sections measured by CMS in electron and muon final
states~\cite{EWK-PAS}.
Similarly, the  normalization of the \ccbar
background relative to  $\PW$+jets  is taken from the  \ttbar cross section  measured by CMS in the dilepton channel~\cite{top}. The measured values of these cross sections are summarized below:
\begin{eqnarray}
\sigma(pp \to {\PW}X) \times \mathrm{B}(\PW \to \ell\nu) &=& 9.951 \pm
0.073~(\mathrm{stat}) \pm 0.280~(\mathrm{syst}) \pm
1.095~(\mathrm{lum})~ \mathrm{nb}~,~ \nonumber \\
\sigma(\Pp\Pp \to {\cPZ}X) \times \mathrm{B}(\cPZ \to \ell\ell) &=&  0.931 \pm 0.026~(\mathrm{stat}) \pm 0.023~(\mathrm{syst}) \pm 0.102~(\mathrm{lum})~ \mathrm{nb}~,~ \\
\sigma(\Pp\Pp \to \ttbar) &=& 194 \pm 72~(\mathrm{stat}) \pm 24~(\mathrm{syst}) \pm 21~(\mathrm{lum})~ \mathrm{pb}~.~\nonumber
\end{eqnarray}

\begin{figure*}[htbp]
\begin{center}
\includegraphics[width=0.49\textwidth]{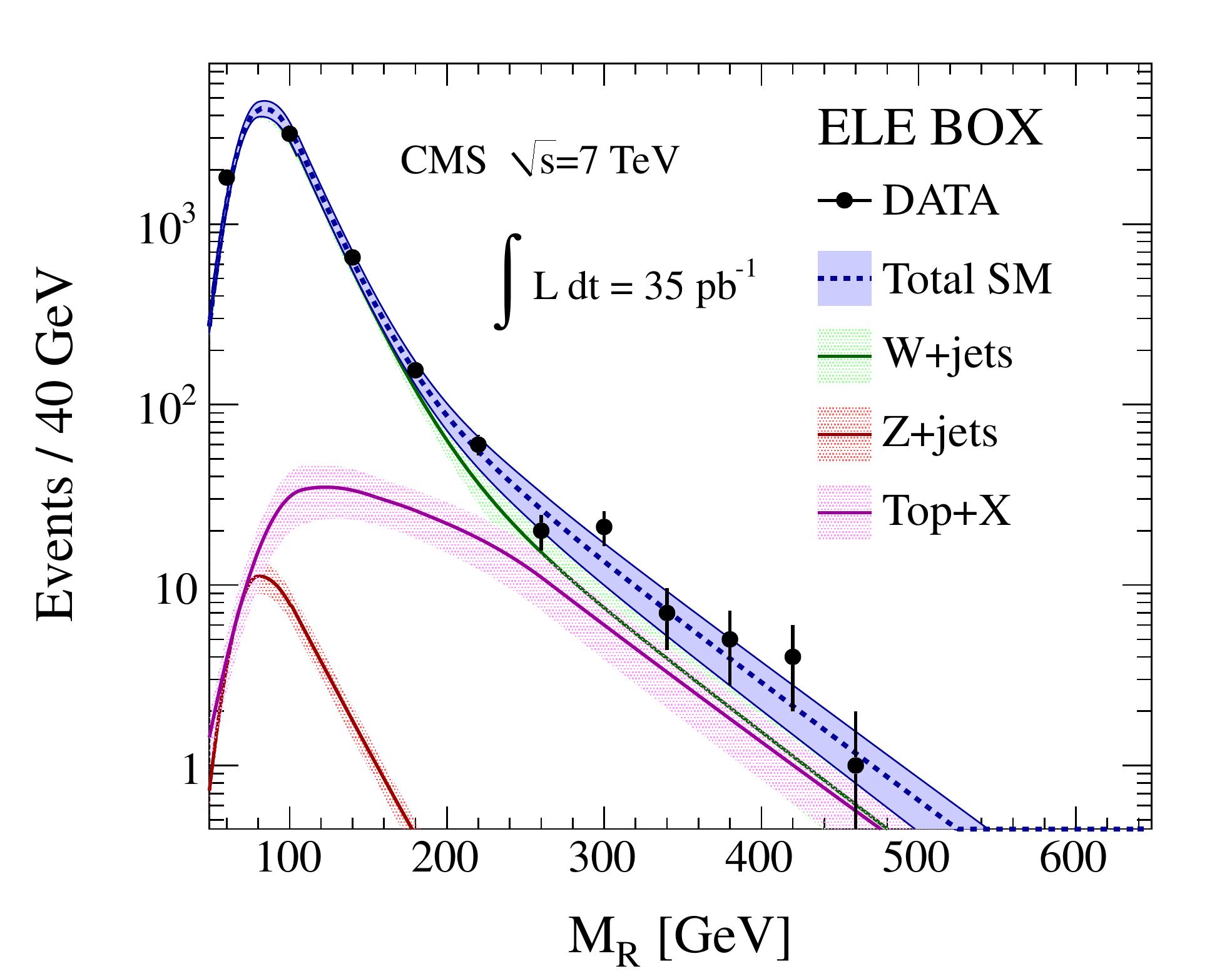}
\includegraphics[width=0.49\textwidth]{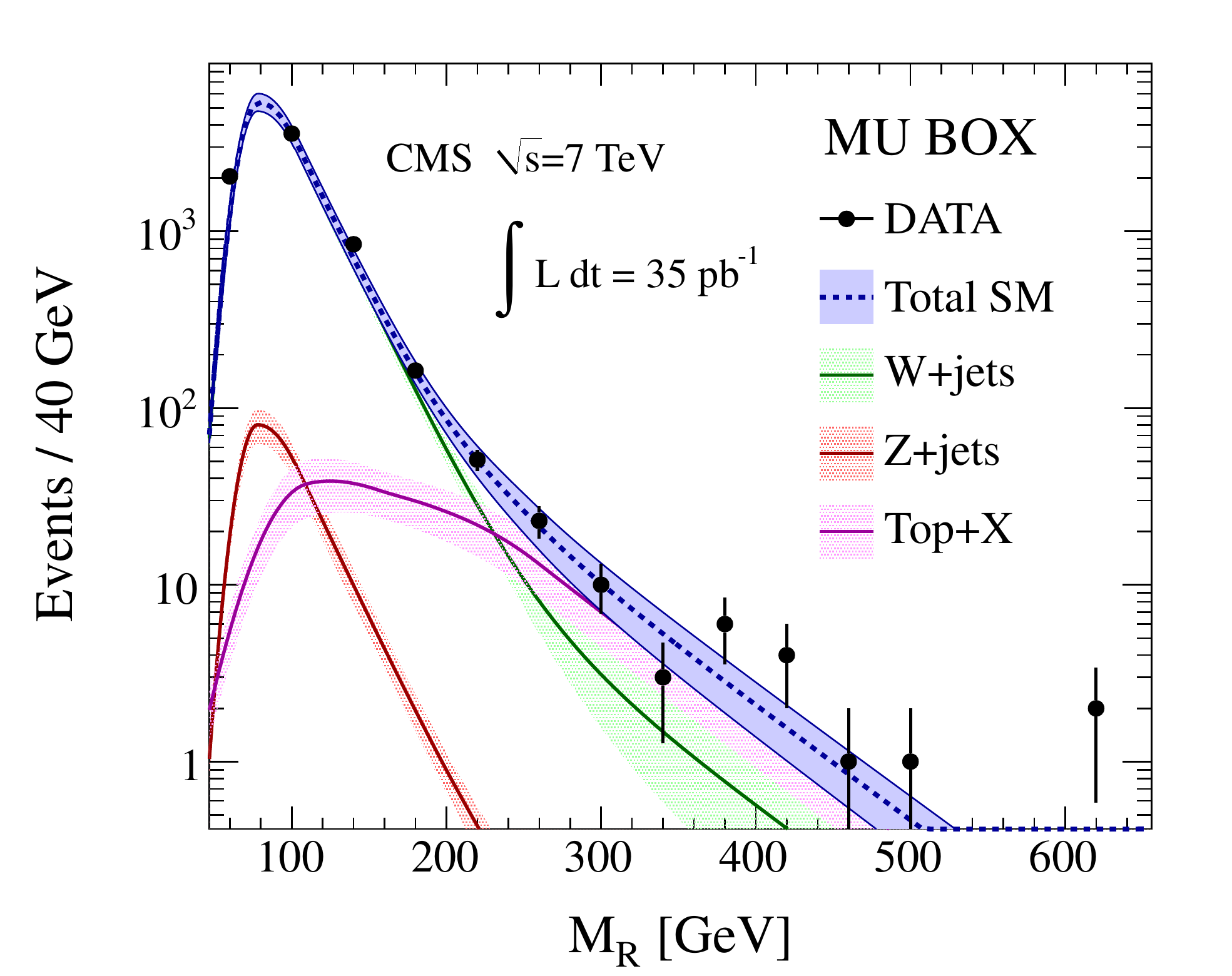}
\caption{The $M_R$ distributions with $R > 0.45$  in the
  ELE (left) and MU (right) boxes for data (points) and backgrounds
  (curves). The bands show the uncertainties of the background predictions.}
\label{fig:ELEMUBOX}
\end{center}
\end{figure*}

For an $R > 0.45$ threshold the QCD background is virtually
eliminated.  The region 125~$< M_{R}< 175\GeV$ where the QCD contribution is
negligible and the $\PW(\ell\nu)$+jets component is dominant is used to
fix the overall normalization of the total background prediction.
The final background prediction in the ELE and MU boxes for $R > 0.45$ is shown in Fig.~\ref{fig:ELEMUBOX}.

The number of events with $M_{R}>500\GeV$ observed in data and the
corresponding number of predicted background events are given in
Table~\ref{tab:ELEMUBOX} for the ELE and MU boxes.
Agreement between the predicted and observed yields is found.  The
$p$-value of the measurement in the MU box is 0.1, given the predicted
background (with its statistical and systematic uncertainties) and the
observed number of
events. 
A summary of the uncertainties entering the background measurements is
presented in Table~\ref{tab:LEPSYS}.
\begin{table}[ht!]
\caption{The number of predicted background events in the  ELE and MU
  boxes for $R>$0.45 and $M_{R}>500\GeV$ and the number of events
  observed in data. \label{tab:ELEMUBOX}}
\smallskip
\centering
\begin{tabular}{|c|c|c|}
\hline
 & Predicted  & Observed \\
\hline
\hline
ELE box &  0.63 $\pm$ 0.23 & 0 \\
\hline
MU box  &  0.51 $\pm$ 0.20 & 3 \\
\hline
\end{tabular}
\end{table}
\begin{table*}[ht!]
  \caption{Summary of the uncertainties on the background predictions
    for  the ELE and MU boxes and their relative magnitudes. The range
    in the Monte Carlo uncertainties is owing to the different
    statistical precisions of the simulated background samples.\label{tab:LEPSYS}}
\smallskip
\centering
\begin{tabular}{|c||c|c|c|}
\hline
Parameter  & Description & Relative magnitude \\
\hline
\hline
Slope parameter $a$ &  systematic bias from correlations in fits & 5\% \\
\hline
Slope parameter $b$ &  systematic bias from correlations in fits & 10\% \\
\hline
Slope parameter $a$ &  uncertainty from Monte Carlo & 1--10\% \\
\hline
Slope parameter $b$ &  uncertainty from Monte Carlo & 1--10\% \\
\hline
$\rho(a)^{\mathrm{data/MC}}$  &   data fit & 3\% \\
\hline
$\rho(b)^{\mathrm{data/MC}}$  &   data fit  & 3\% \\
\hline
Normalization  &  systematic+statistical component & 3--8\% \\
\hline
$f^{\PW}$  &  extracted from fit ($\PW$ only) & 30\% \\
\hline
$PW/\ttbar$ cross section ratio  & CMS measurements (top only) & 40\% \\
\hline
$\PW/\cPZ$ cross section ratio  & CMS measurements ($\cPZ$ only) & 19\% \\
\hline
\end{tabular}
\end{table*}

\subsection{Hadronic box background predictions\label{sec:yesyoucan}}

The procedure for estimating the total background predictions in the
hadronic box is summarized as follows:
\begin{itemize}
\item Construct the non-QCD  background shapes in $M_{R}$ using measured values
  of $a$ and $b$ from simulated events, applying correction
  factors derived from data control samples, and taking into account the $H_T$ trigger turn-on
  efficiency.
\item Set the relative normalizations of the $\PW$+jets, $\cPZ$+jets, and $\cPqt$+X
  backgrounds using the relevant inclusive cross section measurements from
  CMS (Eq. 10).
\item Set the overall normalization by measuring the event yields in
  the lepton boxes, corrected for lepton reconstruction and
  identification efficiencies. The shapes and normalizations of all the
  non-QCD backgrounds are now fixed.
\item The shape of the QCD background is extracted, as described in
  Section ~\ref{sec:qcd}, and its normalization in the HAD box is
  determined from a fit to the low-$M_{R}$ region, as described below.
\end{itemize}

\begin{figure}[htbp]
\begin{center}
\includegraphics[width=0.85\columnwidth]{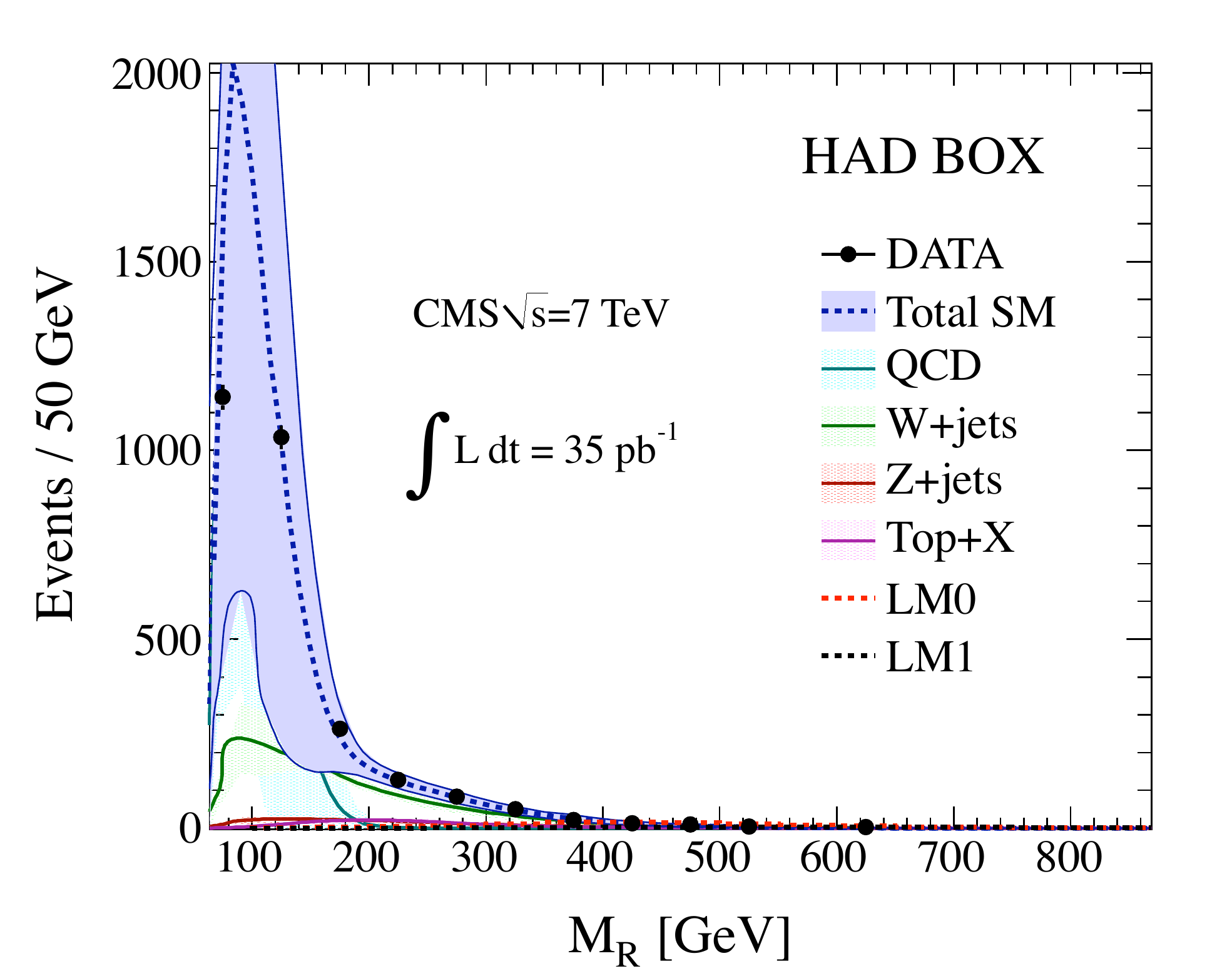}
\includegraphics[width=0.85\columnwidth]{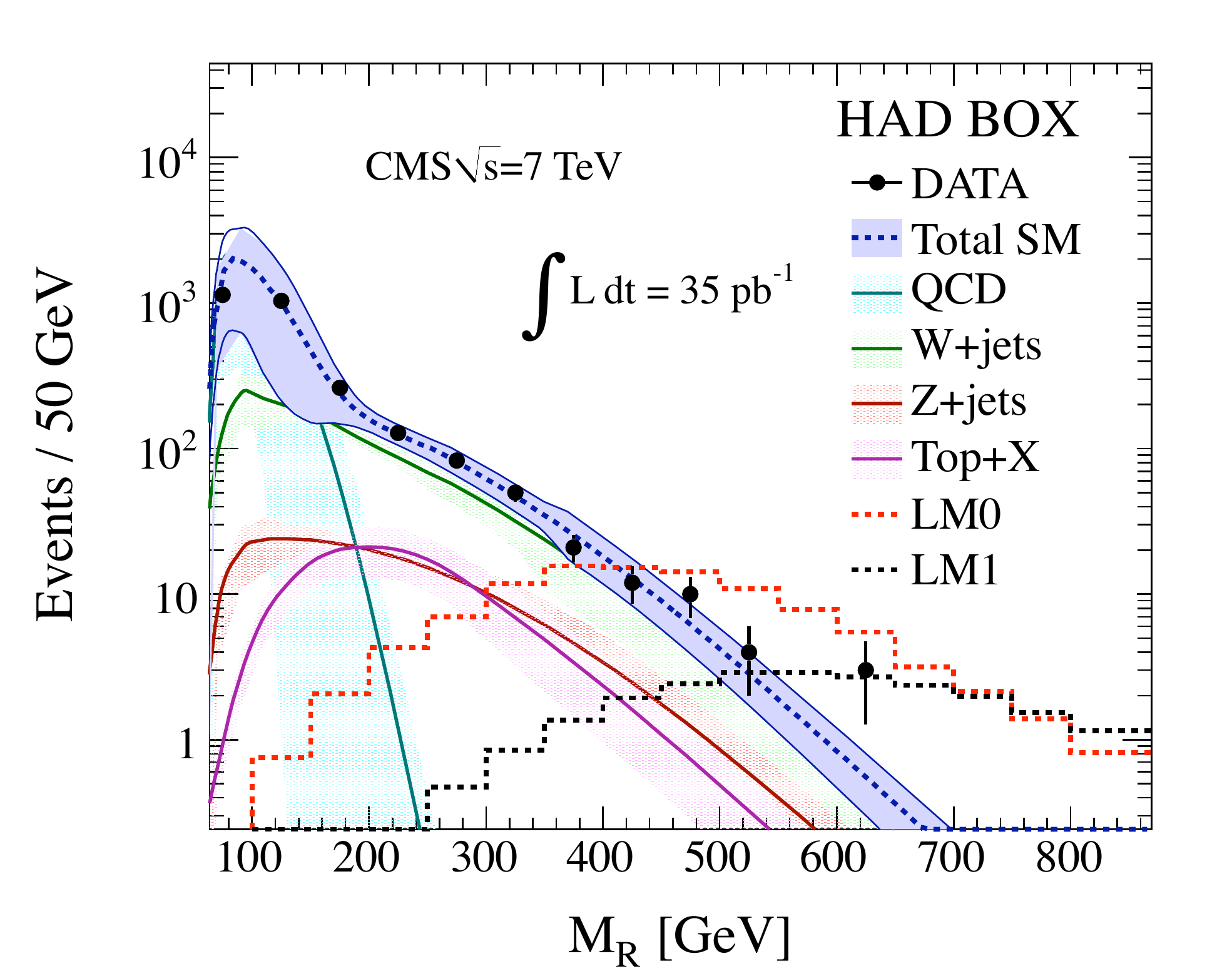}
\caption{The $M_R$ distributions with $R > 0.5$ in the HAD box for
  data (points) and backgrounds (curves) on (top) linear and (bottom)
  logarithmic scales.  The bands show the uncertainties of the
  background predictions. The corresponding distributions for SUSY
  benchmark models LM1 \cite{PTDR2} with $M_\Delta = 597\GeV$ and LM0
  \cite{alphaT} with $M_\Delta = 400\GeV$ are overlaid. }
\label{fig:HADBOX}
\end{center}
\end{figure}

The final hadronic box background prediction is calculated  from a binned
likelihood fit of the total background shape to the data in the
interval 80~$<M_{R}<400\GeV$ with all background normalizations and
shapes fixed, except for the following free parameters: i) the $H_T$
trigger turn-on shapes, ii) $f^{\PW}$ as introduced in
Section~\ref{bg-prop}, and iii) the overall normalization of the QCD
background.  A set of pseudo-experiments is used to test the overall
fit for coverage of the various floated parameters and for systematic
biases.  A 2\% systematic uncertainty is assigned to the high-$M_{R}$
background prediction that encapsulates systematic effects related to
the fitting procedure.  Figure~\ref{fig:HADBOX} shows the final hadronic box
background predictions with all uncertainties on this prediction
included for $R > 0.5$.  The observed $M_{R}$ distribution is consistent with
the predicted one over the entire $M_{R}$ range. The predicted and
observed background yields in the high-$M_{R}$ region are summarized
in Table~\ref{table:PRED}. A summary of the uncertainties entering
these background predictions is listed in Table~\ref{tab:HADSYS}. A
larger $R$ requirement is used in the HAD box analysis due to the
larger background.

\begin{table}[ht!]
\caption{Predicted and observed yields for $M_{R }$$>500\GeV$  with $R > 0.5$ in the HAD box.
\label{table:PRED}}
\smallskip
\centering
\begin{tabular}{|c||c|
c|}
\hline
$M_{R}$   &  Predicted & Observed \\
\hline
\hline
\hline
$M_{R} > 500\GeV$ &  5.5 $\pm$ 1.4 & 7 \\
\hline
\hline
\end{tabular}
\end{table}
\begin{table*}[ht!]
\caption{Summary of uncertainties entering the background predictions for the HAD box.
\label{tab:HADSYS}}
\smallskip
\centering
\begin{tabular}{|c|c|c|}
\hline
Parameter  & Description & Relative magnitude \\
\hline
\hline
Slope parameter $a$ &  systematic bias from correlations in fits & 5\% \\
\hline
Slope parameter $b$ &  systematic bias from correlations in fits & 10\% \\
\hline
Slope parameter $a$ &  uncertainty from Monte Carlo & 1--10\% \\
\hline
Slope parameter $b$ &  uncertainty from Monte Carlo & 1--10\% \\
\hline
$\rho(a)^{\mathrm{data/MC}}$  &  data fit & 3\% \\
\hline
$\rho(b)^{\mathrm{data/MC}}$  & data fit & 3\% \\
\hline
Normalization  &  systematic+statistical component & 8\% \\
\hline
Trigger parameters  & systematic from fit pseudo-experiments & 2\% \\
\hline
$f^{\PW}$  &  extracted from fit ($\PW$ only) & 13\% \\
\hline
$W/\ttbar$ cross section ratio  & CMS measurements (top only) & 40\% \\
\hline
$\PW/\cPZ$ cross section ratio  & CMS measurements ($\cPZ$ only) & 19\% \\
\hline
\end{tabular}
\end{table*}
\section{Limits in the CMSSM Parameter Space}
Having observed no significant excess of events beyond the SM
expectations, we extract a model-independent 95\% confidence level (CL) limit on
the number of signal events. This limit is then
interpreted in the parameter spaces of SUSY models.

The likelihood for the number of observed events $n$ is modeled as a
Poisson function, given the sum of the number of signal events ($s$)
and the number of background events. A posterior probability
density function $P(s)$ for the signal yield is derived using Bayes theorem,
assuming a flat prior for the signal and a log-normal prior for the
background.

The model-independent upper limit is derived by integrating the
posterior probability density function between 0 and $s^*$ so that
$\int_0^{s^*}P(s)ds=0.95$.  The observed upper limit in the hadronic
box is $s^{*}=8.4$ (expected limit 7.2 $\pm$ 2.7); in the muon box
$s^{*}=6.3$ (expected limit 3.5 $\pm$ 1.1); and in the electron box
$s^{*}=2.9$ (expected limit 3.6 $\pm$ 1.1).  For 10\% of the
pseudo-experiments in the muon box the expected limit is higher than
the observed.  The stability of the result was studied with different
choices of the signal prior. In particular, using the reference priors
derived with the methods described in Ref.~\cite{refprior}, the
observed upper limits in the hadronic, muon, and electron boxes
are 8.0, 5.3, and 2.9, respectively.

The results can be interpreted in the context of the CMSSM, which is a
truncation of the full SUSY parameter space motivated by the minimal
supergravity framework for spontaneous soft breaking of
supersymmetry. In the CMSSM the soft breaking parameters are reduced
to five: three mass parameters $m_0$, $m_{1/2}$, and $A_0$ being,
respectively, a universal scalar mass, a universal gaugino mass, and a
universal trilinear scalar coupling, as well as $\tan\beta$, the
ratio of the up-type and down-type Higgs vacuum expectation values,
and the sign of the supersymmetric Higgs mass parameter
$\mu$. Scanning over these parameters yields models which, while not
entirely representative of the complete SUSY parameter space, vary
widely in their superpartner spectra and thus in the dominant
production channels and decay chains.

The upper limits are projected onto the ($m_0$, $m_{1/2}$) plane by
comparing them with the predicted yields, and excluding any model if
$s(m_{0},m_{1/2})>s^{*}$.  The systematic uncertainty on the signal
yield (coming from the uncertainty on the luminosity, the selection
efficiency, and the theoretical uncertainty associated with the cross
section calculation) is modeled according to a log-normal prior. The
uncertainty on the selection efficiency includes the effect of
jet energy scale (JES) corrections, parton distribution function (PDF)
uncertainties~\cite{Bourilkov:2006cj}, and the description of
initial-state radiation (ISR). All the effects are summed in
quadrature as shown in Table~\ref{tab:syst}. The JES, ISR, and PDF
uncertainties are relatively small owing to the insensitivity of the
signal $R$ and $M_R$ distributions to these effects.

\begin{table}[ht!]
  \caption{Summary of the systematic uncertainties on the signal yield
    and totals for each of the event boxes. For the CMSSM scan the NLO signal cross section uncertainty is included.  \label{tab:syst}}
\centering
\smallskip
\begin{tabular}{|lccc|}
\hline
box & MU & ELE & HAD \\\hline
\multicolumn{4}{|c|}{Experiment}\\\hline
JES & 1\% & 1\% & 1\% \\\hline
Data/MC $\epsilon$& 6\% & 6\% & 6\% \\\hline
$\mathcal{L}$\cite{lumi-moriond} & 4\% & 4\% & 4\% \\
\hline\hline
\multicolumn{4}{|c|}{Theory}\\\hline
ISR & 1\% & 1\% & 0.5\% \\\hline
PDF & 3--6\% & 3--6\% & 3--6\% \\ \hline
Subtotal & 8--9\% & 8--9\% & 8--9\% \\
\hline\hline
\multicolumn{4}{|c|}{CMSSM}\\\hline
NLO  & 16--18\% & 16--18\% & 16--18\% \\\hline
Total & 17--19\% & 17--19\% & 17--19\% \\
\hline\hline
\end{tabular}
\end{table}

\begin{figure}[htbp]
\begin{center}
\includegraphics[width=0.85\columnwidth]{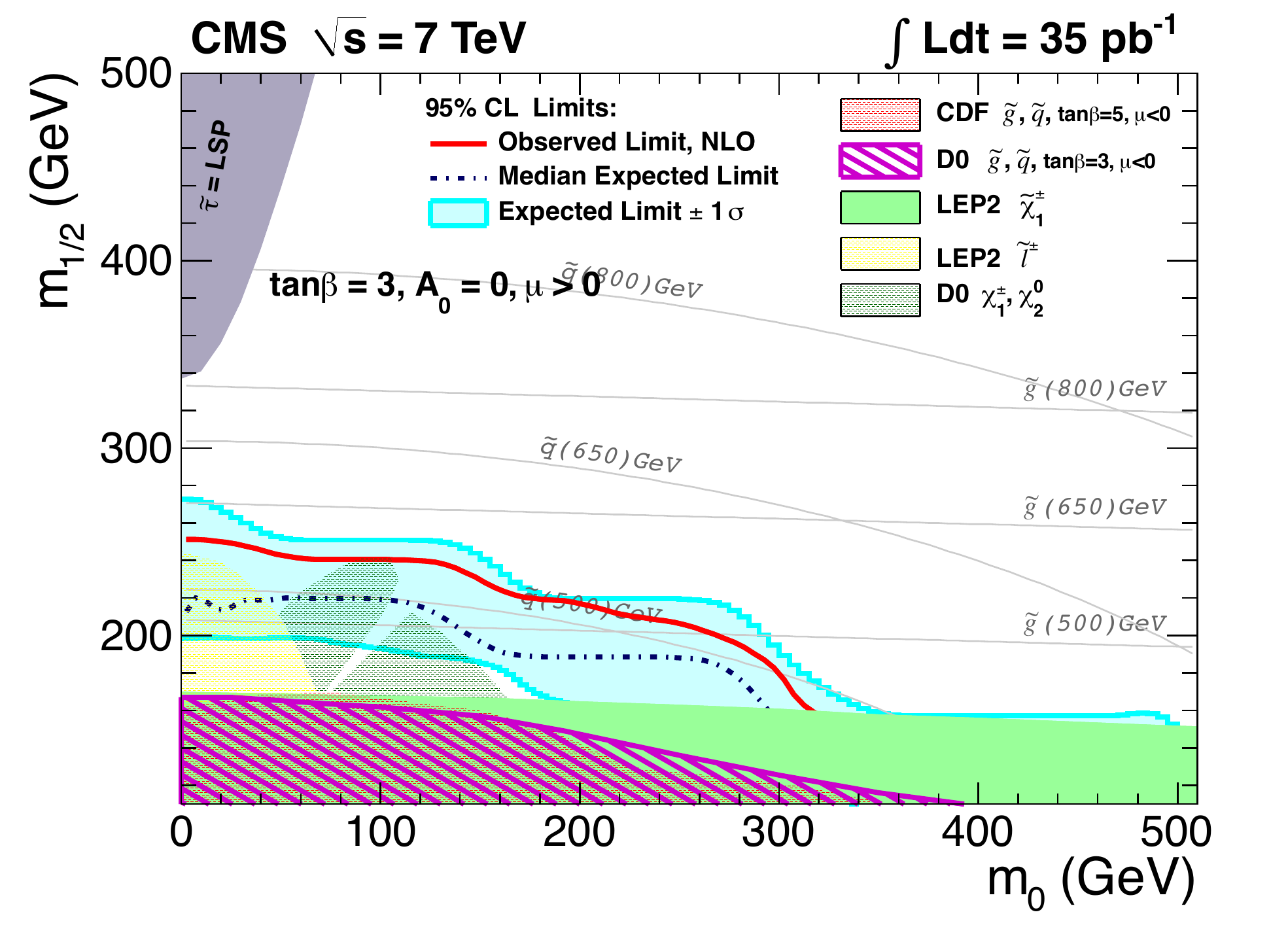}
\caption{Observed (solid curve) and expected (dot-dashed curve) 95\% CL
  limits in the ($m_{0}$, $m_{1/2}$) CMSSM plane with $\tan\beta=3$,
  $A_{0} = 0$, $\operatorname{sgn}(\mu) = +1$ from  the ELE box selection ($R >
  0.45$, $M_{R} > 500\GeV$). The $\pm$ one standard deviation
  equivalent variations in the uncertainties are shown as a band
  around the expected limits. The area labeled $\tilde{\tau}$=LSP
  is the region of the parameter space where the LSP is a
  $\tilde{\tau}$ and not the lightest neutralino.}
\label{fig:ELE_LIMIT}
\end{center}
\end{figure}

\begin{figure}[htbp]
\begin{center}
\includegraphics[width=0.85\columnwidth]{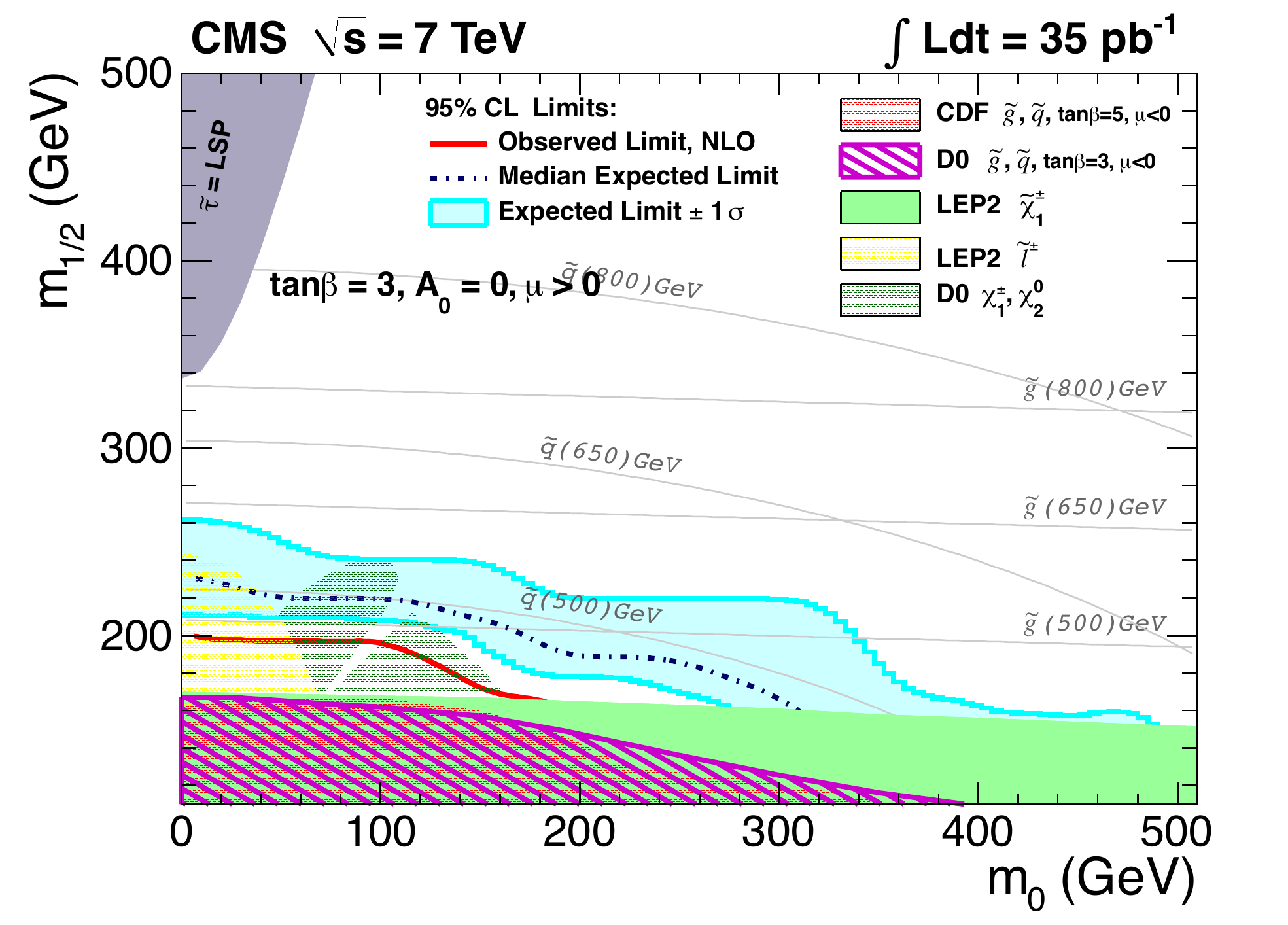}
\caption{Observed (solid curve) and expected (dot-dashed curve) 95\% CL
  limits in the ($m_{0}$, $m_{1/2}$) CMSSM plane with $\tan\beta=3$,
  $A_{0} = 0$, $\operatorname{sgn}(\mu) = +1$ from  the MU box selection ($R >
  0.45$, $M_{R} > 500\GeV$). The $\pm$ one standard deviation
  equivalent variations in the uncertainties are shown as a band
  around the expected limits. }
\label{fig:test}
\end{center}
\end{figure}

\begin{figure}[htbp]
\begin{center}
\includegraphics[width=0.85\columnwidth]{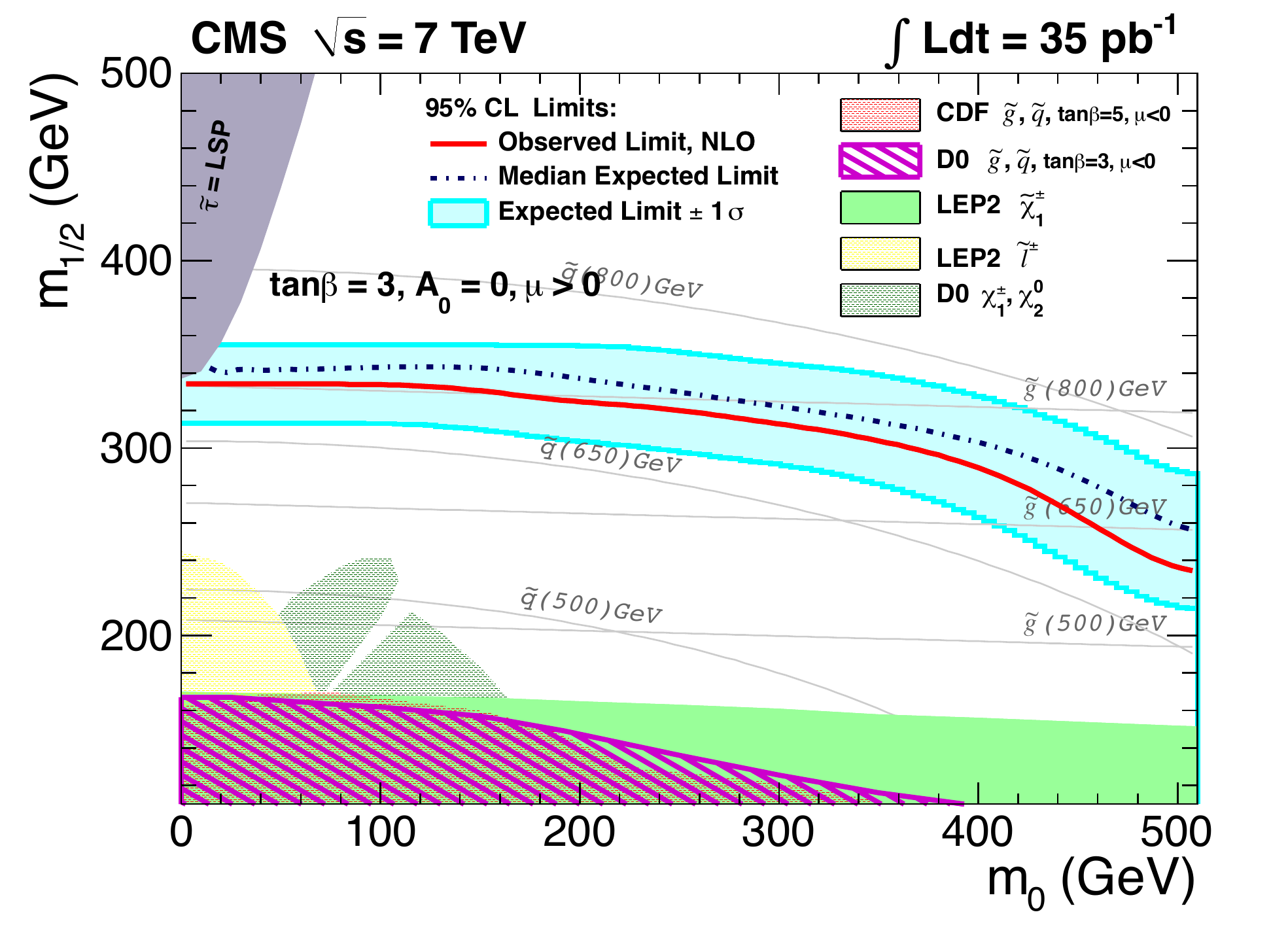}
\caption{Observed (solid curve) and expected (dot-dashed curve) 95\% CL
  limits in the ($m_{0}$, $m_{1/2}$) CMSSM plane with $\tan\beta=3$,
  $A_{0} = 0$, $\operatorname{sgn}(\mu) = +1$ from  the HAD box selection ($R >
  0.5$, $M_{R} > 500\GeV$). The $\pm$ one standard deviation
  equivalent variations in the uncertainties are shown as a band
  around the expected limits.  }
\label{fig:HAD_LIMIT}
\end{center}
\end{figure}

\begin{figure}[h!]
\begin{center}
\includegraphics[width=0.85\columnwidth]{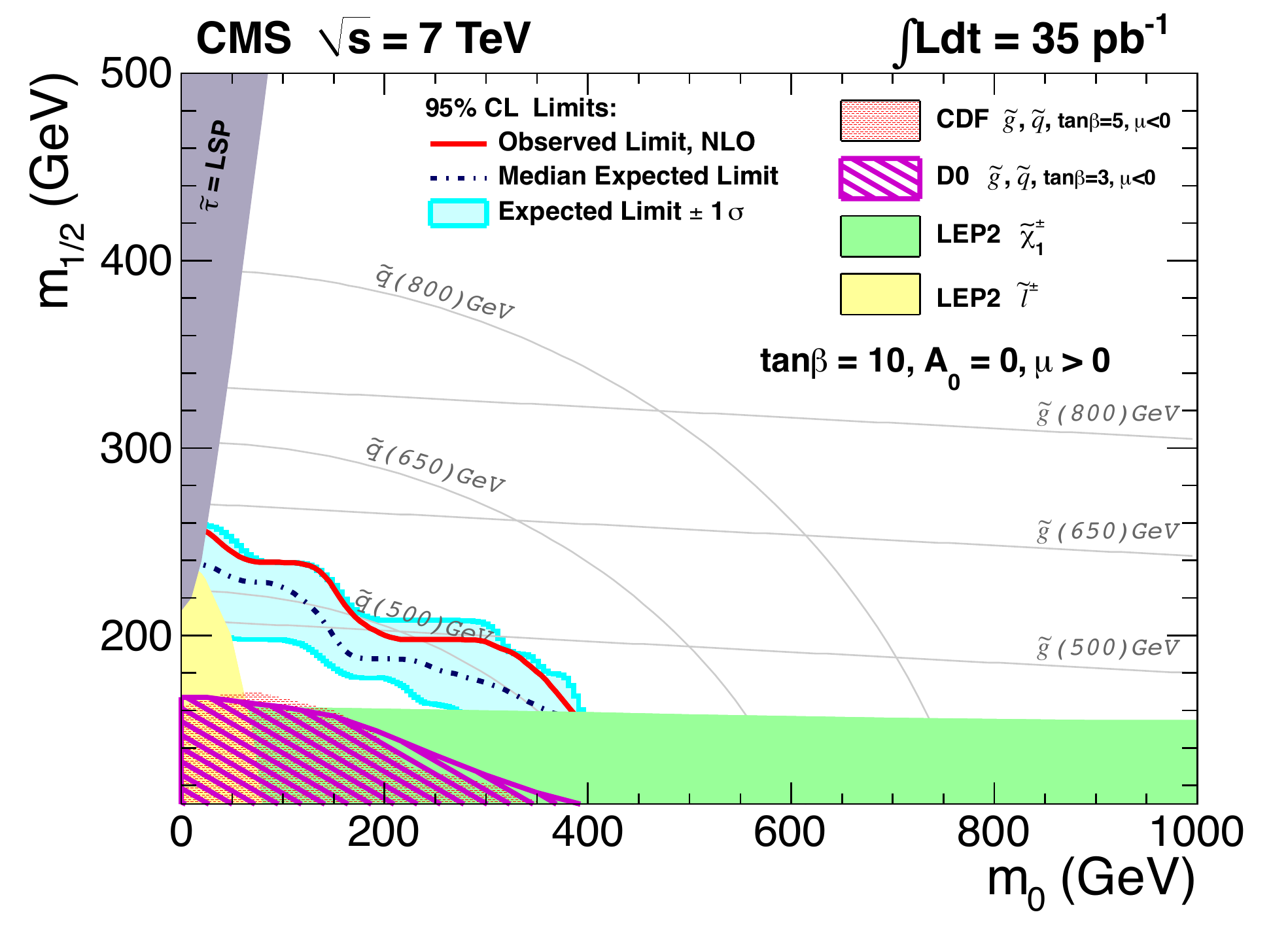}
\caption{Observed (solid curve) and expected (dot-dashed curve) 95\% CL
  limits in the ($m_{0}$, $m_{1/2}$) CMSSM plane with $\tan\beta=10$,
  $A_{0} = 0$, $\operatorname{sgn}(\mu) = +1$ from  the ELE box selection ($R >
  0.45$, $M_{R} > 500\GeV$). The $\pm$ one standard deviation
  equivalent variations in the uncertainties are shown as a band
  around the expected limits.  }
\label{fig:ELE_LIMIT10}
\end{center}
\end{figure}

\begin{figure}[h!]
\begin{center}
\includegraphics[width=0.85\columnwidth]{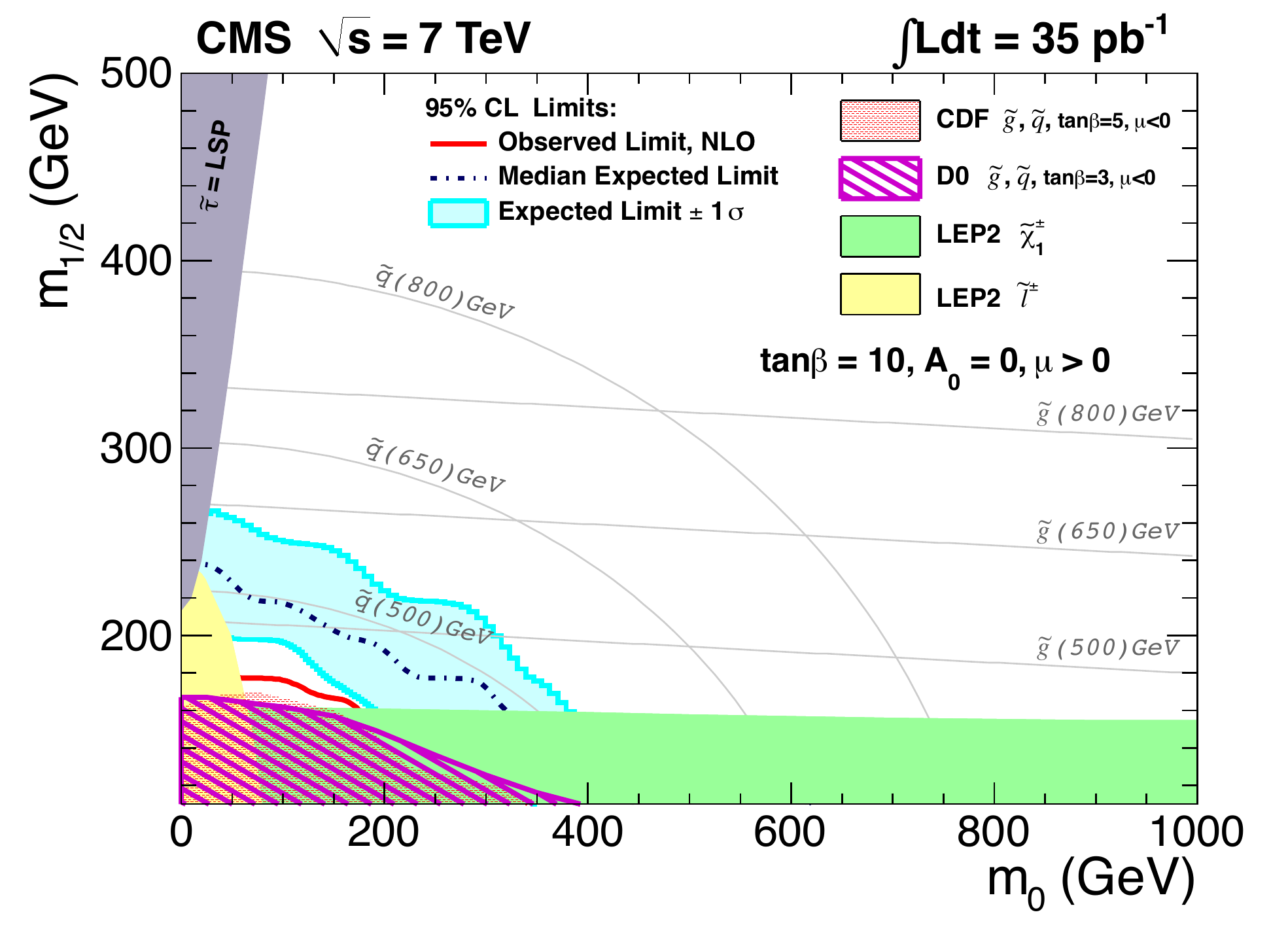}
\caption{Observed (solid curve) and expected (dot-dashed curve) 95\% CL
  limits in the ($m_{0}$, $m_{1/2}$) CMSSM plane with $\tan\beta=10$,
  $A_{0} = 0$, $\operatorname{sgn}(\mu) = +1$ from  the MU box selection ($R >
  0.45$, $M_{R} > 500\GeV$). The $\pm$ one standard deviation
  equivalent variations in the uncertainties are shown as a band
  around the expected limits.  }
\label{fig:MU_LIMIT10}
\end{center}
\end{figure}

\begin{figure}[h!]
\begin{center}
\includegraphics[width=0.85\columnwidth]{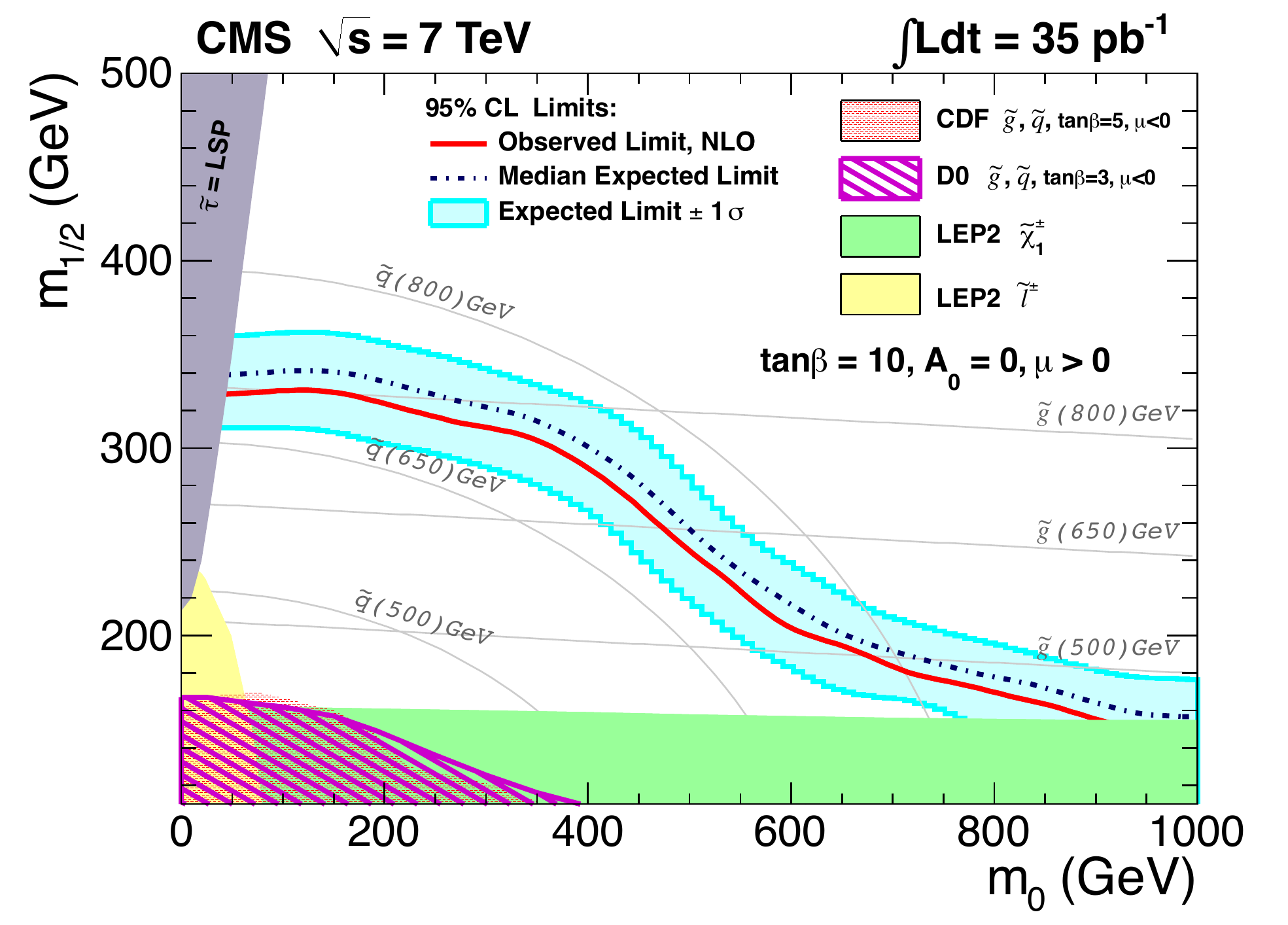}
\caption{Observed (solid curve) and expected (dot-dashed curve) 95\% CL
  limits in the ($m_{0}$, $m_{1/2}$) CMSSM plane with $\tan\beta=10$,
  $A_{0} = 0$, $\operatorname{sgn}(\mu) = +1$ from  the HAD box selection ($R >
  0.5$, $M_{R} > 500\GeV$). The $\pm$ one standard deviation
  equivalent variations in the uncertainties are shown as a band
  around the expected limits.  }
\label{fig:HAD_LIMIT10}
\end{center}
\end{figure}

\begin{figure}[h!]
\begin{center}
\includegraphics[width=0.85\columnwidth]{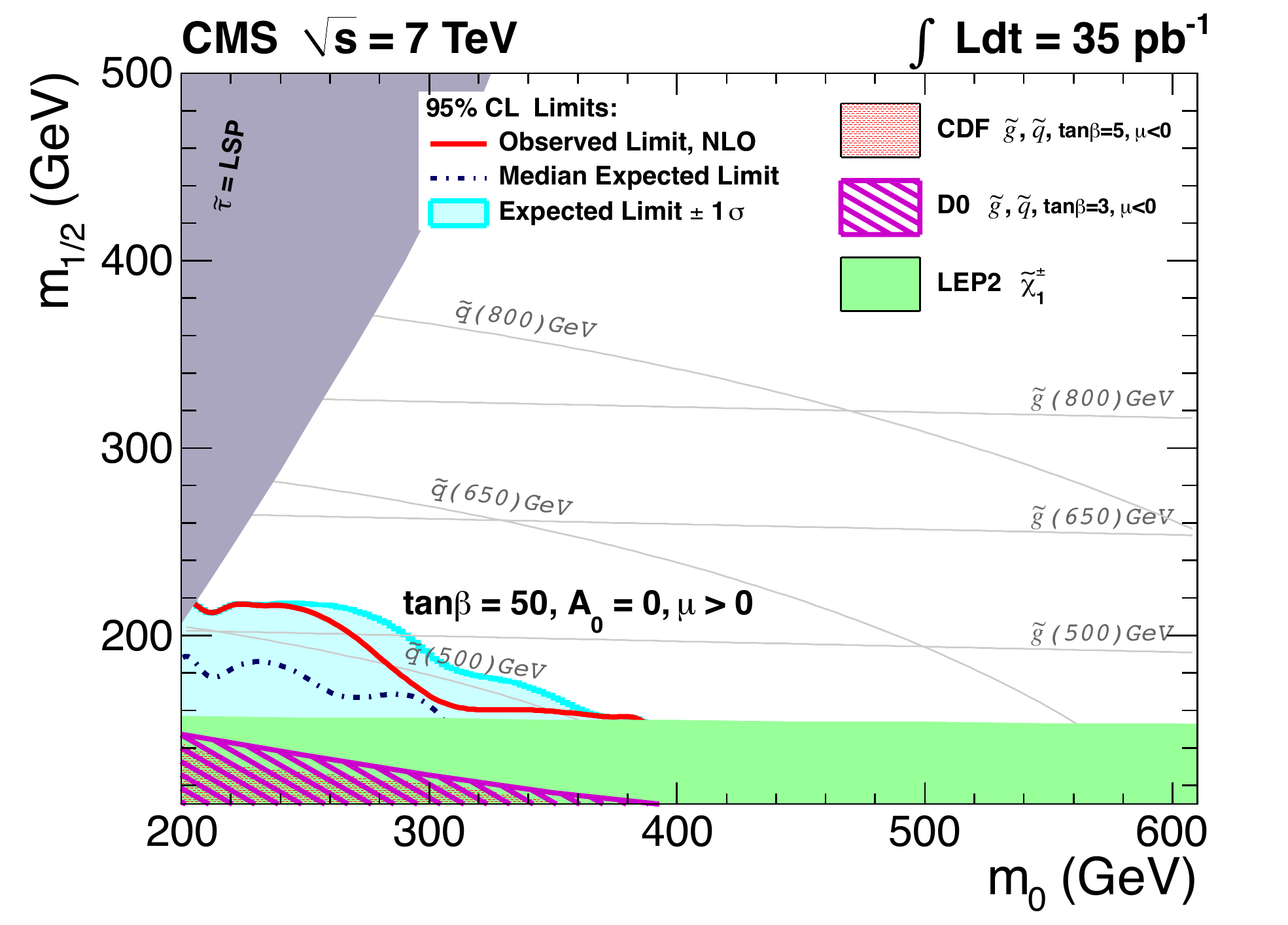}
\caption{Observed (solid curve) and expected (dot-dashed curve) 95\% CL
  limits in the ($m_{0}$, $m_{1/2}$) CMSSM plane with $\tan\beta=50$,
  $A_{0} = 0$, $\operatorname{sgn}(\mu) = +1$ from  the ELE box selection ($R >
  0.45$, $M_{R} > 500\GeV$). The $\pm$ one standard deviation
  equivalent variations in the uncertainties are shown as a band
  around the expected limits.  }
\label{fig:ELE_LIMIT50}
\end{center}
\end{figure}

\begin{figure}[h!]
\begin{center}
\includegraphics[width=0.85\columnwidth]{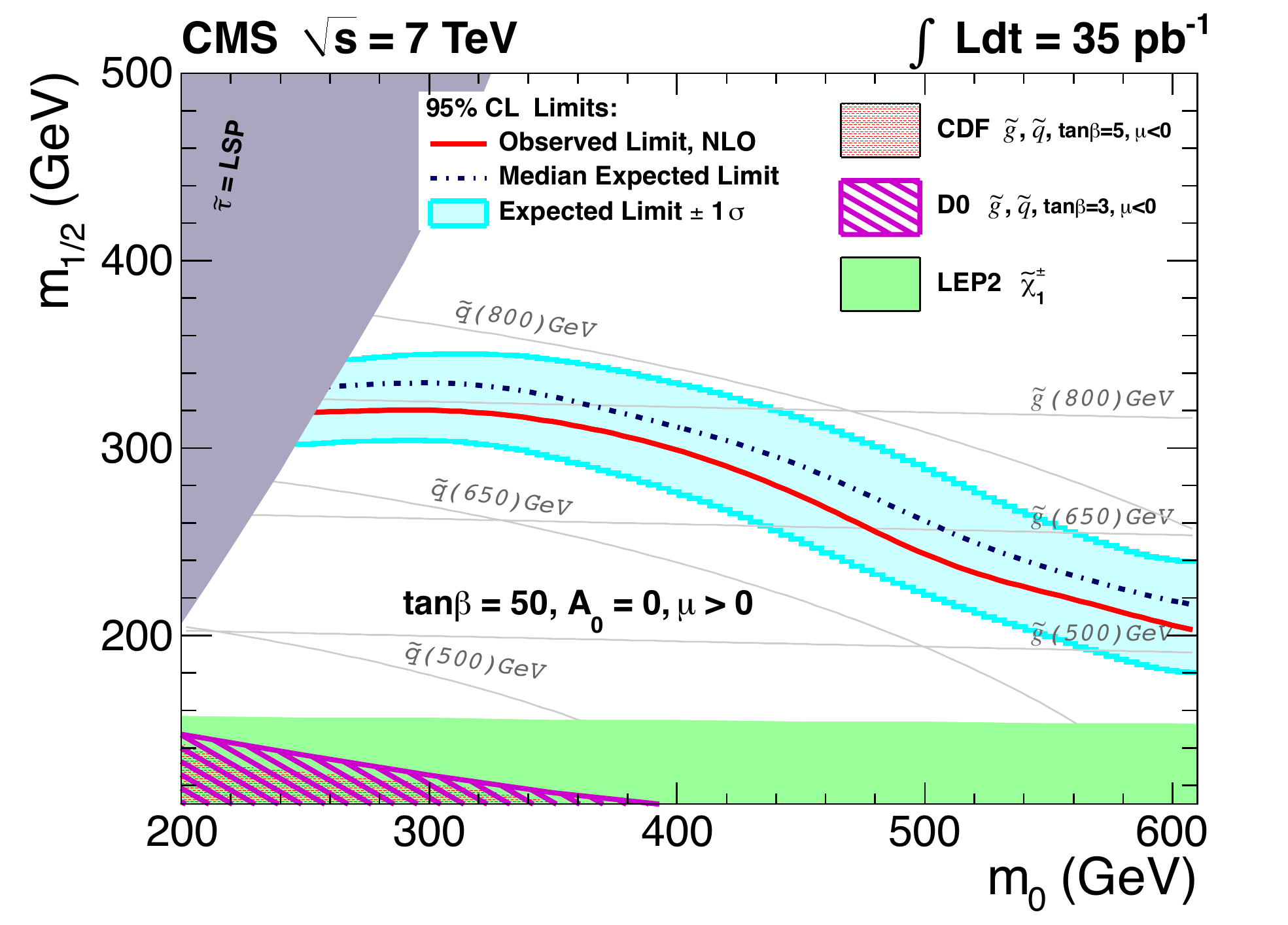}
\caption{Observed (solid curve) and expected (dot-dashed curve) 95\% L
  limits in the ($m_{0}$, $m_{1/2}$) CMSSM plane with $\tan\beta=50$,
  $A_{0} = 0$, $\operatorname{sgn}(\mu) = +1$ from  the HAD box selection ($R >
  0.5$, $M_{R} > 500\GeV$). The $\pm$ one standard deviation
  equivalent variations in the uncertainties are shown as a band
  around the expected limits.  }
\label{fig:HAD_LIMIT50}
\end{center}
\end{figure}

The observed limits from the ELE, MU, and HAD boxes are shown in
Figs.~\ref{fig:ELE_LIMIT},~\ref{fig:test}, and~\ref{fig:HAD_LIMIT},
respectively, in the CMSSM
($m_{0}$, $m_{1/2}$) plane for the values $\tan\beta = 3$, $A_{0} = 0$,
$\operatorname{sgn}(\mu) = +1$,
 together with the 68\% probability band for the
expected limits, obtained by applying the same procedure to an ensemble of
background-only  pseudo-experiments. The band is computed
around the median of the limit distribution.  Observed limits are also shown
in Figs.~\ref{fig:ELE_LIMIT10}
--\ref{fig:HAD_LIMIT10} in the CMSSM ($m_{0}$, $m_{1/2}$) plane for the
values $\tan\beta = 10$, $A_{0} = 0$,
$\operatorname{sgn}(\mu) = +1$, and in
Figs.~\ref{fig:ELE_LIMIT50}--\ref{fig:HAD_LIMIT50} for the values
$\tan\beta = 50$, $A_{0} = 0$,
$\operatorname{sgn}(\mu) = +1$.

\begin{figure}[h!]
\begin{center}
\includegraphics[width=0.7\columnwidth]{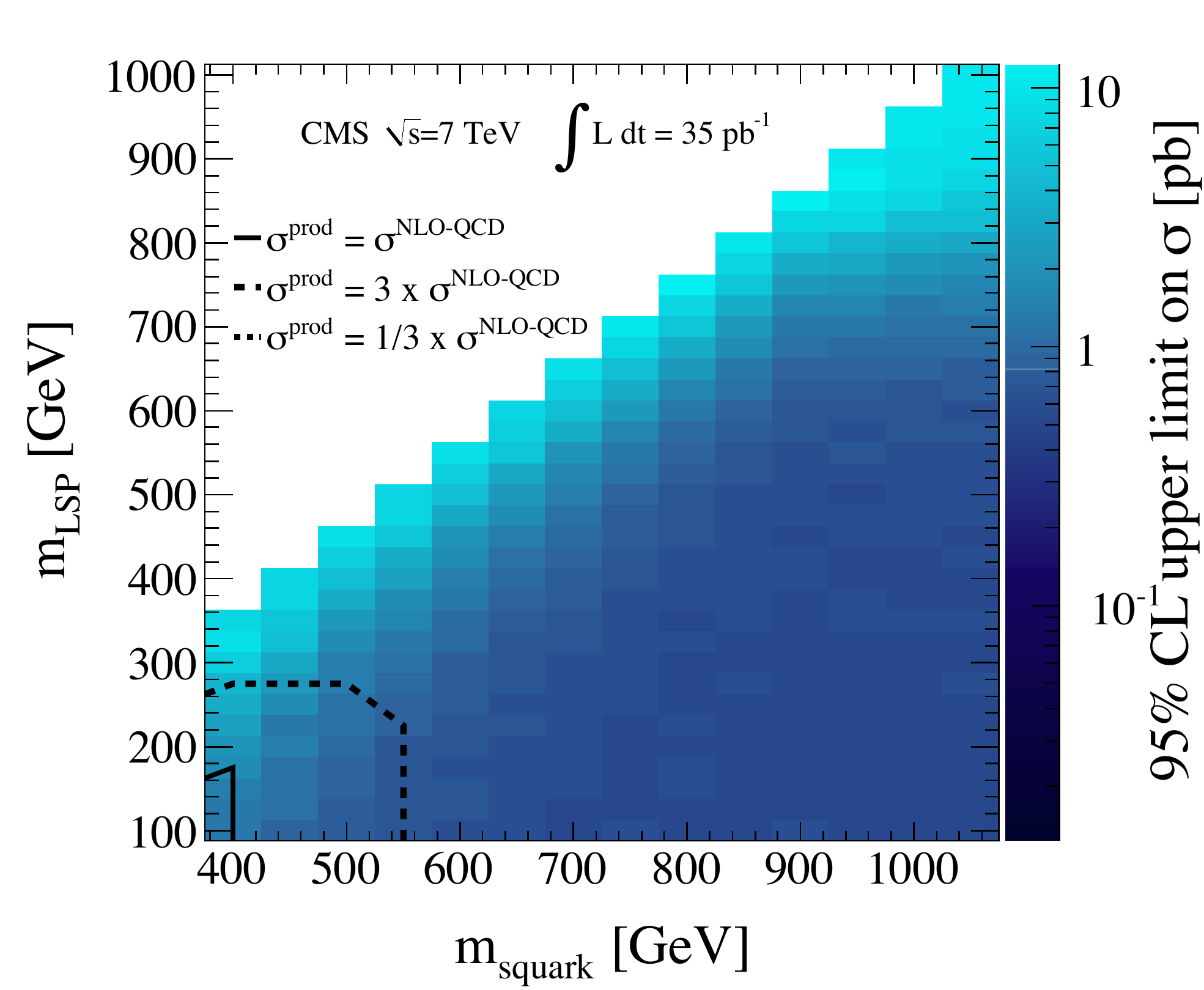}
\includegraphics[width=0.7\columnwidth]{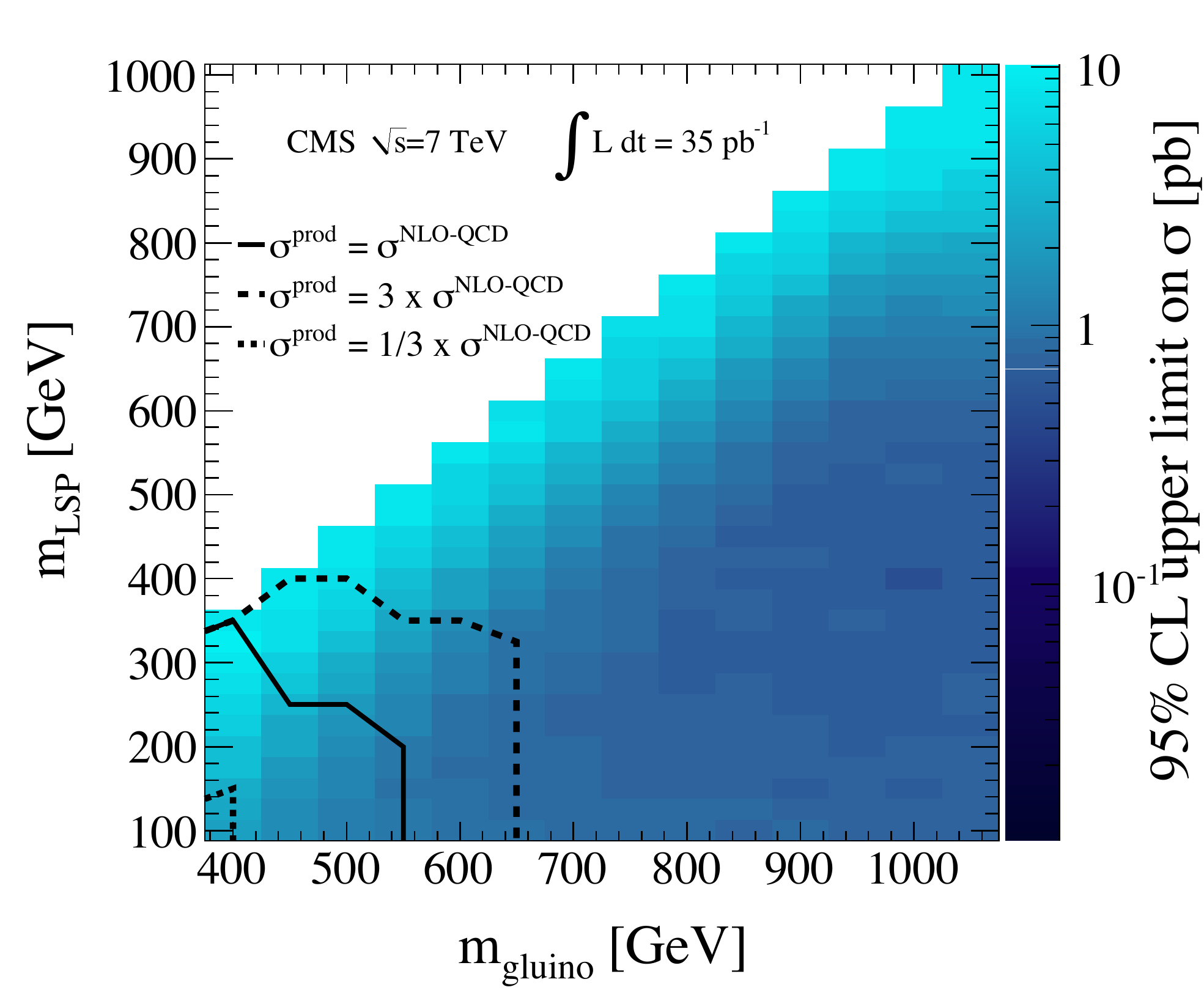}
\caption{Upper limits on two simplified models: di-squark production (top)
  resulting in a 2-jet + \ETm final state and di-gluino (lower)
  production resulting in a 4-jet + \ETm final state. The
  shade scale indicates the value of the cross section excluded at 95\%
  CL  for each value of $m_{\text{LSP}}$ and $m_{\text{gluino}}$
  or $m_{\text{squark}}$. The solid and dashed contours indicate the
  95\% CL limits assuming the NLO cross section and its variations
  up and down by a factor of three.}
\label{fig:SMS}
\end{center}
\end{figure}

Figure ~\ref{fig:SMS} shows the same result in terms of 95\% CL upper
limits on the cross section as a function of the physical masses for two benchmark
simplified models~\cite{Alwall-1,Alwall-2,Sanjay,RA2}: four-flavor
squark pair production  and gluino pair production. In
the former, each squark decays to one quark and the LSP, resulting in
final states with two jets and missing transverse energy, while in the
latter each gluino decays directly to two light quarks and the LSP,
giving events with four jets and missing transverse energy.




\section{Summary}
We performed a  search for squarks and gluinos using a data
sample of 35\pbinv integrated luminosity from pp collisions at
$\sqrt{s} = 7\TeV$, recorded by the CMS detector at the LHC.  The
kinematic consistency of the selected events was tested against the
hypothesis of heavy particle pair production using the dimensionless
razor variable $R$ related to the missing transverse energy \ETm,
and $M_R$, an event-by-event indicator of the heavy particle
mass scale.  We used events with large $R$ and high $M_R$ in inclusive
topologies.

The search relied on predictions of the SM backgrounds determined from
data samples dominated by SM processes.  No significant excess over
the background expectations was observed, and model-independent upper
limits on the numbers of signal events were calculated.
The results were presented in the ($m_0$, $m_{1/2}$) CMSSM parameter
space. For simplified models the results were given as  limits on the
production cross sections as a function of the squark, gluino, and LSP
masses.

These results demonstrate the strengths of the razor analysis
approach; the simple exponential behavior of the various SM
backgrounds when described in terms of the razor variables is useful
in suppressing these backgrounds and in making reliable estimates from
data of the background residuals in the signal regions. Hence, the
razor method provides an additional powerful probe in searching for
physics beyond the SM at the LHC.

\section*{Acknowledgments}

\hyphenation{Bundes-ministerium Forschungs-gemeinschaft Forschungs-zentren} We wish to congratulate our colleagues in the CERN accelerator departments for the excellent performance of the LHC machine. We thank the technical and administrative staff at CERN and other CMS institutes. This work was supported by the Austrian Federal Ministry of Science and Research; the Belgian Fonds de la Recherche Scientifique, and Fonds voor Wetenschappelijk Onderzoek; the Brazilian Funding Agencies (CNPq, CAPES, FAPERJ, and FAPESP); the Bulgarian Ministry of Education and Science; CERN; the Chinese Academy of Sciences, Ministry of Science and Technology, and National Natural Science Foundation of China; the Colombian Funding Agency (COLCIENCIAS); the Croatian Ministry of Science, Education and Sport; the Research Promotion Foundation, Cyprus; the Estonian Academy of Sciences and NICPB; the Academy of Finland, Finnish Ministry of Education and Culture, and Helsinki Institute of Physics; the Institut National de Physique Nucl\'eaire et de Physique des Particules~/~CNRS, and Commissariat \`a l'\'Energie Atomique et aux \'Energies Alternatives~/~CEA, France; the Bundesministerium f\"ur Bildung und Forschung, Deutsche Forschungsgemeinschaft, and Helmholtz-Gemeinschaft Deutscher Forschungszentren, Germany; the General Secretariat for Research and Technology, Greece; the National Scientific Research Foundation, and National Office for Research and Technology, Hungary; the Department of Atomic Energy and the Department of Science and Technology, India; the Institute for Studies in Theoretical Physics and Mathematics, Iran; the Science Foundation, Ireland; the Istituto Nazionale di Fisica Nucleare, Italy; the Korean Ministry of Education, Science and Technology and the World Class University program of NRF, Korea; the Lithuanian Academy of Sciences; the Mexican Funding Agencies (CINVESTAV, CONACYT, SEP, and UASLP-FAI); the Ministry of Science and Innovation, New Zealand; the Pakistan Atomic Energy Commission; the State Commission for Scientific Research, Poland; the Funda\c{c}\~ao para a Ci\^encia e a Tecnologia, Portugal; JINR (Armenia, Belarus, Georgia, Ukraine, Uzbekistan); the Ministry of Science and Technologies of the Russian Federation, and Russian Ministry of Atomic Energy; the Ministry of Science and Technological Development of Serbia; the Ministerio de Ciencia e Innovaci\'on, and Programa Consolider-Ingenio 2010, Spain; the Swiss Funding Agencies (ETH Board, ETH Zurich, PSI, SNF, UniZH, Canton Zurich, and SER); the National Science Council, Taipei; the Scientific and Technical Research Council of Turkey, and Turkish Atomic Energy Authority; the Science and Technology Facilities Council, UK; the US Department of Energy, and the US National Science Foundation.

Individuals have received support from the Marie-Curie programme and the European Research Council (European Union); the Leventis Foundation; the A. P. Sloan Foundation; the Alexander von Humboldt Foundation; the Associazione per lo Sviluppo Scientifico e Tecnologico del Piemonte (Italy); the Belgian Federal Science Policy Office; the Fonds pour la Formation \`a la Recherche dans l'Industrie et dans l'Agriculture (FRIA-Belgium); the Agentschap voor Innovatie door Wetenschap en Technologie (IWT-Belgium); and the Council of Science and Industrial Research, India.
\cleardoublepage
\bibliography{auto_generated}

\cleardoublepage \appendix\section{The CMS Collaboration \label{app:collab}}\begin{sloppypar}\hyphenpenalty=5000\widowpenalty=500\clubpenalty=5000\textbf{Yerevan Physics Institute,  Yerevan,  Armenia}\\*[0pt]
S.~Chatrchyan, V.~Khachatryan, A.M.~Sirunyan, A.~Tumasyan
\vskip\cmsinstskip
\textbf{Institut f\"{u}r Hochenergiephysik der OeAW,  Wien,  Austria}\\*[0pt]
W.~Adam, T.~Bergauer, M.~Dragicevic, J.~Er\"{o}, C.~Fabjan, M.~Friedl, R.~Fr\"{u}hwirth, V.M.~Ghete, J.~Hammer\cmsAuthorMark{1}, S.~H\"{a}nsel, M.~Hoch, N.~H\"{o}rmann, J.~Hrubec, M.~Jeitler, W.~Kiesenhofer, M.~Krammer, D.~Liko, I.~Mikulec, M.~Pernicka, B.~Rahbaran, H.~Rohringer, R.~Sch\"{o}fbeck, J.~Strauss, A.~Taurok, F.~Teischinger, P.~Wagner, W.~Waltenberger, G.~Walzel, E.~Widl, C.-E.~Wulz
% ...
\vskip\cmsinstskip
% ...
1:~~Also at CERN, European Organization for Nuclear Research, Geneva, Switzerland\\
% ...

\end{sloppypar}
\end{document}